\documentclass[a4paper,fleqn]{cas-sc}

\usepackage[authoryear]{natbib}
\usepackage{xcolor}

\def\tsc#1{\csdef{#1}{\textsc{\lowercase{#1}}\xspace}}
\tsc{WGM}
\tsc{QE}

\begin{document}
\let\WriteBookmarks\relax
\def\floatpagepagefraction{1}
\def\textpagefraction{.001}

\shorttitle{Accretion of Uranus and Neptune}    

\shortauthors{Esteves et al.}  

\title [mode = title]{Accretion of Uranus and Neptune: confronting different giant impact scenarios}  



%

\author[1,2]{Leandro Esteves}

\cormark[1]


\ead{leandro.esteves@unesp.br}

\credit{Conceptualization of the study, Methodology, Software programming, Simulation performing and Writing}

\affiliation[1]{organization={UNESP, São Paulo State University, Grupo de Dinâmica Orbital e Planetologia},
            city={Guaratinguetá},
            postcode={12516-410}, 
            state={São Paulo},
            country={Brazil}}
\affiliation[2]{organization={Université Paris Cité, Institut de Physique du Globe de Paris},
            city={Paris},
            postcode={F-75005}, 
            state={île de France},
            country={France}}

\author[3]{André Izidoro}


\ead{izidoro.costa@gmail.com}

\credit{Conceptualization of the study, Methodology, Software programming and Writing}

\affiliation[3]{organization={Rice University, Department of Earth, Environmental and Planetary Sciences},
            city={Houston},
            postcode={TX 77005}, 
            state={Texas},
            country={USA}}

\cortext[1]{Corresponding author}

\fntext[1]{}

\author[1]{Othon C. Winter}

\credit{Conceptualization of the study, Methodology and Writing}


\begin{abstract}
The origins of Uranus and Neptune are not fully understood. Their inclined rotation axes – obliquities – suggest that they experienced giant impacts during their formation histories. Simulations modeling their accretion from giant impacts among $\sim$5 Earth masses planetary embryos -- with roughly unity impactors' mass ratios -- have been able to broadly match their current masses, final mass ratio, and obliquity. However, due to angular momentum conservation, planets produced in these impacts tend to rotate too fast, compared to Uranus and Neptune. One potential solution for this problem consists of invoking instead collisions of objects with large mass ratios (e.g. a proto-Uranus with 13~M$_{\oplus}$ and an \textcolor{black}{embryo} of 1~M$_{\oplus}$). Smooth-particle hydrodynamics simulations show that in this scenario final planets tend to have rotation periods more consistent with those of Uranus and Neptune. Here we performed a large suite of N-body numerical simulations modelling the formation of Uranus and Neptune to compare these different dynamical views. Our simulations start with a population of protoplanets and account for the effects of type-I migration, inclination and eccentricity tidal damping. Our results show that although scenarios allowing for large impactors' mass ratio favour slower rotating planets, the probability of occurring collisions in these specific simulations is significantly low. This is because gas tidal damping is relatively less efficient for low-mass embryos ($\lesssim$1~M$_{\oplus}$) and, consequently, such objects are mostly scattered by more massive objects ($\sim$13~M$_{\oplus}$) instead of colliding with them. Altogether, our results show that the probability of broadly matching the masses, mass ratio, and rotation periods of Uranus and Neptune in these two competing formation scenarios is broadly similar, within a factor of $\sim$2, with overall probabilities of the order of $\sim$0.1-1\%.
\end{abstract}


\begin{highlights}
\item Both large and small impactor scenarios remain viable for the accretion of Uranus and Neptune from a planet formation perspective.
\item The probability of collisions in simulations with small impactors ($\lesssim$1~M$_{\oplus}$) is significantly low but the rotation periods of the resulting planets peak around the present-day values of Uranus and Neptune.
\item Large impactor scenarios exhibit a higher probability of collisions but most planets have excess angular momentum.
\end{highlights}


\maketitle

\section{Introduction}
The origins of Uranus and Neptune remain an open question. Uranus and Neptune are intermediate-sized planets with approximately 14.5 Earth masses (M$_{\oplus}$) and 17.1~M$_{\oplus}$, respectively. Interior models suggest that volatile elements such as H and He make up about 2 to 3~M$_{\oplus}$ of their total mass, while the remaining mass is in form of heavier elements~\citep{helledetal11, nettelmannetal13, helledetal20}. The abundance of volatile elements suggests that Uranus and Neptune probably formed in the presence of the Solar Nebula and accreted their H and He content from the gas disk itself~\citep{safronov72,pollacketal96}. Some models even suggest that they could be cores of giant planets that failed to undergo runaway gas accretion and to acquire large gaseous envelopes~\citep{dodsonrobinsonbodenheimer10,helledbodenheimer14,helled23}. The formation of planetary objects with masses of Uranus and Neptune at their current location is generally a difficult problem in planet formation models.

Planet formation models suggest that centimeter- and millimeter-sized dust particles, known as pebbles, sediment in the midplane of the protoplanetary disk. Upon decoupling from the gas motion, these pebbles experience a drag force that generates a radial drift towards the star~\citep{adachietal76,nakagawaetal86}. The most recent planetesimal formation models indicate that substructures in the gas disk such as pressure bumps, efficiently trap pebbles in specific regions, establishing the conditions required for planetesimal formation within the natal disk~\citep{youdingoodman05,johansenetal07,lyraetal08,drazkowskaetal16,guilerasandor17,drazkowskadullemond18,morbidelli20,izidoroetal22b,lauetal22,sandoretal24,lauetal24}. The subsequent growth of planetesimals into planetary cores may take place via planetesimal  or pebble accretion, or a combination of these processes.

In the case of Uranus and Neptune, it is quite unlikely that they formed at their current location exclusively by planetesimal accretion~ \citep{wetherillstewart89,kokuboida98,chambers01}. As planets grow embedded in a disk of planetesimals, they open gaps in the radial distribution of planetesimals and experience a dramatic reduction in their own growth rate~\citep{tanakaida97,levisonetal10}. The formation of Uranus and Neptune  via planetesimal accretion exclusively leads to very long timescales that generally exceeds the typical life of gaseous protoplanetary disks~\citep{levisonstewart01,thommesetal03,levisonmorbidelli07}. 

Pebble accretion, on the other hand, may lead to growth of planets of several Earth masses in relatively much shorter timescales \citep{lambrechtsetal12, johansenetal15, levisonetal15, bitschetal15b, lambrechtsetal19}.  Pebbles may be accreted by a growing planet when they drift in the disk due to the effects of gas drag~\citep{johansenetal07, ormelklahr10, johansenlambrechts17}. In the presence of a sufficient flow of pebbles, protoplanets of up to several Earth masses my grow in very short timescales (e.g. 100~kyr) of even at large orbital separations, as those of Uranus and Neptune~\citep{lambrechtsetal12,vallettahelled22}.

Regardless of their very formation process, and how pebble or planetesimal accretion  contribute to their final growth, it is quite likely that Uranus and Neptune also experienced at least one giant impact each early in their history. This hypothesis is supported by different lines of evidence. The first evidence is their significant obliquities (spin-axis tilt), which are approximately 97° and 30°, respectively. \textcolor{black}{Potential scenarios to generate such large obliquities include giant impacts during their formation process~\citep{morbidellietal12}, or alternatively, dynamical processes occurring after planet formation~\citep{bouelaskar10,rogoszinskihamilton21}. The second evidence} is that it is also reasonable to envision that Uranus and Neptune did not reached their final masses purely by pebble accretion. Pebble accretion  eventually stops when planets reach pebble isolation mass~\citep{lambrechtsetal12}, and pebble isolation mass decreases as the disk ages~\citep{bitschetal18}. If Uranus and Neptune formed relatively late compared to Jupiter and Saturn, their pebble isolation in the sun's natal disk mass may have been relatively low \citep[e.g.][]{izidoroetal21}, and potentially lower than their current masses which would require subsequent giant impacts. Finally, additional evidences supporting the giant impact hypothesis includes the presence of regular satellites around Uranus and Neptune's inferred interior structure ~\citep{podolakhelled12,morbidellietal12,kegerreisetal18,kurosakiinutsuka19,reinhardtetal20,wooetal2022,salmoncanup22}. We argue that a combination of pebble and planetesimal accretion and a subsequent phase of giant impacts among Earth-masses (or larger) objects stands as one of the most promising avenue for explaining their origins. 

Furthermore, this envisioned giant impact phase is more likely to have occurred during the gas disk phase rather than after disk dispersal, as suggested by their H-He rich atmospheres. H-He rich atmospheres may be lost due to late (after disk dispersal) giant impacts~\citep{bierstekerschlichting19}. Giant impacts taking place after disk dispersal may imply that these planets may not have the chance to re-accrete their gas envelopes. Secondly, giant impacts between bodies at the orbital distances of Uranus and Neptune after gas disk dispersal are also relatively difficult to occur due to the long orbital period timescales. At such large distances from the sun, planet-planet scattering dominates over accretion. During the gas disk phase, on the other hand, sufficiently massive planets undergo gas-driven type-I planet migration~\citep{goldreichtremaine80,ward86,tanakaetal02} which typically is inwards~\citep{izidoroetal15,izidoroetal15b,piranietal21}. Gas-driven migration may lead to convergence of protoplanets towards specific regions of the disk facilitating close-encounters and eventually giant impacts~\citep{izidoroetal15}.

The most recent simulations of the formation of Uranus and Neptune, modelling their accretion from a population of migrating embryos, assume a population of about 5 to 10 protoplanets of $\sim$3 to 6 M$_{\oplus}$ each ~\citep{jakubiketal12,izidoroetal15} beyond Saturn. Jupiter and Saturn are assumed fully formed and in a resonant configuration, typically the 3:2 mean motion resonance. Protoplanets initially distributed beyond Saturn are allowed to migrate inwards, leading to scattering events and eventually collisions with each other~\citep{izidoroetal15}. Simulations by \citet{izidoroetal15} (hereafter I15) modelling this scenario have been successful in reproducing final planets with masses compatible to those of Uranus and Neptune, and with a large range of obliquities. In the nominal simulation of I15, the best Uranus and Neptune analogues are result of one or two giant impacts each, where colliding bodies have generally mass-ratios of the order of unity.

Recent smoothed particle hydrodynamics (SPH) simulations using the impact data from  \citet{izidoroetal15} have shown that such impacts typically produce planets with final rotation periods too short compared to those of Uranus and Neptune~\citep{chauetal21}. This occurs because impacts involving bodies with similar masses results in a significant transfer of angular momentum in form of rotation to the resulting planet~\citep{kokuboetal07,kegerreisetal18,reinhardtetal20}. Although it is possible that effects such as tidal dissipation, resonant effects~\citep[e.g.][]{cukstewart12}, primordial satellites, and gas-planet interaction can remove some angular momentum from certain systems, it is not clear how efficiently these processes are in the case of Uranus and Neptune. 

The natural competing hypothesis to overcome this issue is to invoke that Uranus and Neptune were tilted instead by giant impacts between objects with large mass ratios (e.g. a proto-Uranus with $13~M_{\oplus}$ and an embryo of $1~M_{\oplus}$). In this case, the transfer of angular momentum during the impact tend to not lead to very short rotation periods for these planets~\citep{reinhardtetal20}. SPH simulations exploring a large set of impact conditions confirm that a high target-to-impactor mass ratio alleviate this issue~\citep[e.g.][]{kegerreisetal18,kurosakiinutsuka19,reinhardtetal20,wooetal2022}. Specifically for Uranus, these simulations suggest that impactors with a mass of $\lesssim$ 3~M$_{\oplus}$ are required to reproduce both the planet's obliquity and the presence of a debris disk massive enough to subsequently form the Uranian satellites system~\citep{reinhardtetal20,wooetal2022}. 

In this paper, we revisit the final phase of the accretion of Uranus and Neptune, following the model of  I15, but assuming a less restrictive range of protoplanet masses from which Uranus and Neptune could have accreted from. We performed a suite of numerical simulations assuming different combination of initial masses and  numbers of protoplanetary embryos. Our simulations model the  migration and dynamical evolution of protoplanets embedded in a gaseous disk. We compared the results of our new simulations with those of previous studies in order to infer which scenario is statistically more likely to account for the formation of Uranus and Neptune within the context of solar system formation and evolution models. 

The paper is organized as follows. In Section \ref{sec:sim}, we describe the simulations and the parameters used in each set of simulations. In Section \ref{sec:results}, we show the fraction of simulations that present the desired collisions for each set. We analyse rotation periods and obliquities of resulting planets, calculate the probabilities and compare the between the scenarios. \textcolor{black}{In Section \ref{sec:discussion} we discuss the caveats of our model.} Finally, in Section \ref{sec:summary}, we summarize the implication of our results.

\section{Simulations}\label{sec:sim}

We performed our N-body simulations using a modified version of the hybrid integrator from the Mercury package~\citep{chambers99}, adapted to model the accretion of Uranus and Neptune as in I15. Our code incorporates artificial forces to mimic the effects of disk-planet tidal interactions by adding the corresponding additional accelerations at each time step. We assume that the protoplanetary disk follows the classical minimum mass solar nebula disk~\citep{massetetal06,walshmorbidelli11}. The gas disk surface density profile is extracted from 2D hydrodynamics simulations~\citep{morbidellicrida07} accounting for  the effects of fully formed Jupiter and Saturn in the disk. We use as input into our N-body code the 1D-azimuthally averaged radial disk profiles. We account for the gas disk's dissipation due to photo-evaporation and viscous accretion onto the Sun by assuming an exponential decay of the disk's gas density over time. We include the effects of type-I migration using the locally isothermal disk approximation to compute the planetary torques~\citep{paardekooperetal11}. We also include inclination and eccentricity damping following the analytic prescriptions from~\citep{tanakaward04,cresswellnelson06,cresswellnelson08}. We neglect the large-scale migration of Jupiter and Saturn in the disk, and we damp their orbital inclinations and eccentricities following the prescription of I15~\textcolor{black}{(based on \cite{cridaetal08})} in order to mimic the effects of the gas disk onto Jupiter and Saturn. We set the eccentricity and inclination damping timescales for Jupiter as $e_j/\dot{e_j} \sim10^4$ yr and $i_j/\dot{i_j} \sim10^5$ yr, respectively. For Saturn we use shorter timescales with $e_s/\dot{e_s} \sim10^3$ yr and $i_s/\dot{i_s} \sim10^4$. To keep these planets in non-migrating orbits we restore the planets' semi-axis value on a timescale of 1~Myr. \textcolor{black}{For our simulations, we use a timestep set to 1/25th of the orbital period of the innermost body. Close encounters are handled using a Bulirsch-Stoer integration scheme. The systems are simulated for up to 3 Myr, when gas disk is assumed to have dissipated.}

{
\begin{figure}
    \centering
    \includegraphics[scale=0.52]{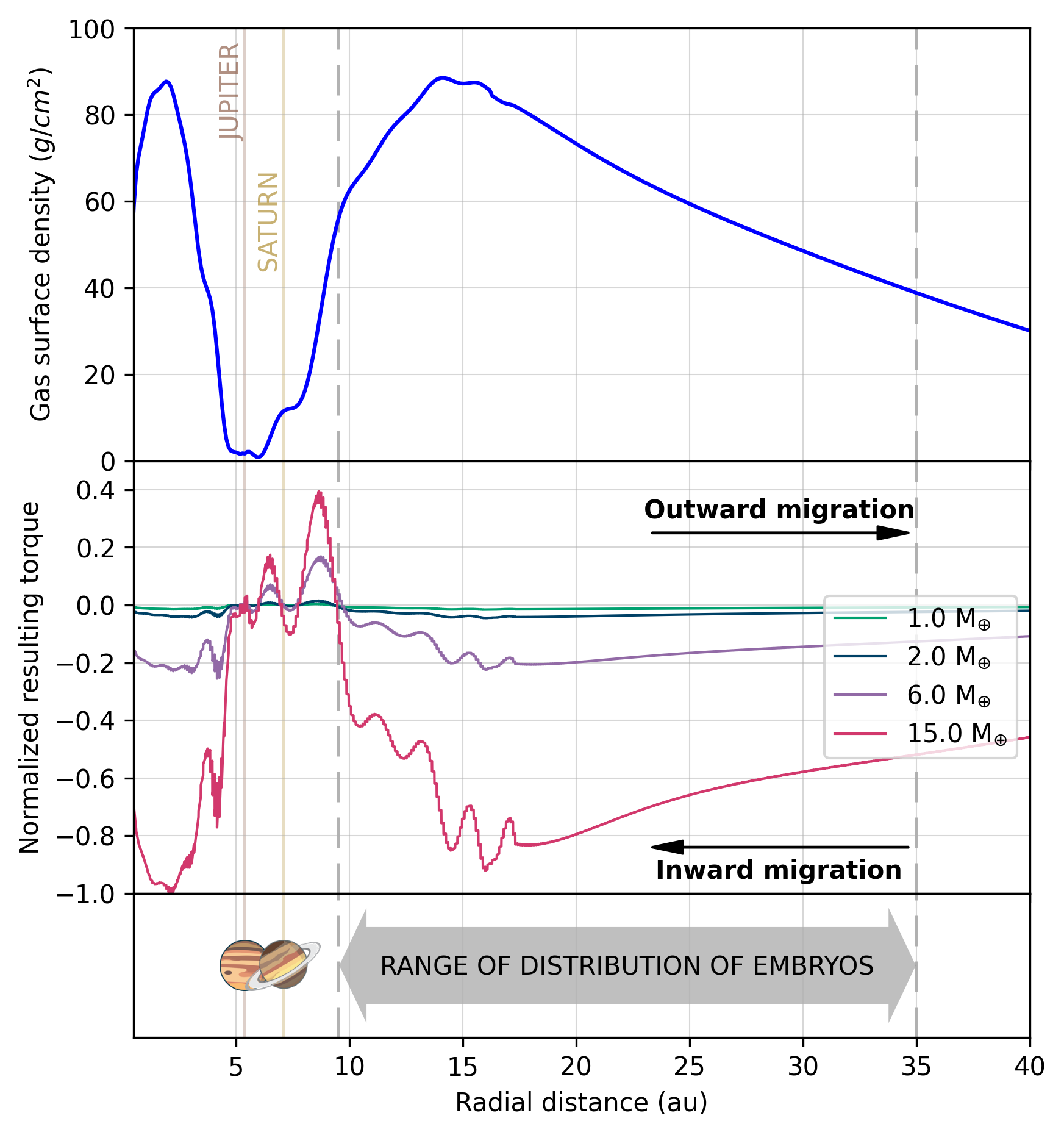}
    \caption{Simulation initial conditions for the protoplanetary disk and the planets.
    \textbf{Top panel:} The gas surface density as a function of radial distance. The blue curve illustrates how the giant planets shape the disk. Coloured lines represent the approximate orbits of Jupiter, Saturn, and the range of distribution for embryos.    
    \textbf{Middle Panel:} Normalized resultant torque. Negative values indicate a reduction in orbital angular momentum leading to migration towards the Sun, while positive value means outward migration. The colour-coding of lines corresponds to bodies with different masses.
    \textbf{Bottom Panel:} Approximate initial positions of Jupiter, Saturn, and the embryos that will give rise to planets analogous to Uranus and Neptune.}
    \label{fig:initialcond}
\end{figure}
}

Figure \ref{fig:initialcond} top panel shows the gas surface density profile assumed in our simulations. This figures shows the locations of Jupiter and Saturn within the disk and the gaps created by these planets~\citep{morbidellicrida07}. The middle panel of Figure \ref{fig:initialcond} shows the torque experiencing by migrating protoplanets of different masses migrating in the disk. As protoplanets approach Saturn the resulting total torque becomes positive which reverses their inward migration. 

As explained before, in I15 simulations, initial planetary embryos distributed beyond Saturn are roughly equal mass bodies (mass ratio of the order of unity) with masses varying from 3 to 6~M$_{\oplus}$. This scenario successfully explains the masses of Uranus and Neptune but leads to giant impacts where final planets generally rotating too fast. In order to explore the formation of Uranus and Neptune via giant impacts involving objects with high mass ratios (e.g. mass-ratio of the order of 10), we selected different initial masses for Uranus' and Neptune's building blocks from those assumed in I15. Our initial population of planetary embryos comes in two classes of objects, the so-called protoplanets (or specifically refereed as proto-Uranus or proto-Neptune) and relatively smaller planetary embryos.

Our simulations start with at least two protoplanets and different number of small embryos (N$_{\textrm{small}}$), with masses set as M$_{\textrm{small}}$. The initial masses of the protoplanets in our simulations is a free parameter and assumed to vary between 11 and 16.5~M$_{\oplus}$, depending on the initial masses of the small embryos. For instance, for a set of simulations where small embryos have initial masses set to M$_{\textrm{small}}$, the proto-Uranus and proto-Neptune start the simulations with masses set to $14 - $M$_{\textrm{small}}$ and $17 - $M$_{\textrm{small}}$, respectively. We choose these specific protoplanets masses such that only a single giant impact is needed to achieve Uranus/Neptune current masses. We performed simulations considering different values for N$_{\textrm{small}}$, as 5, 10, and 20. We also considered four different individual masses for small embryos, as M$_{\textrm{small}}$: 0.5~M$_{\oplus}$, 1~M$_{\oplus}$, 2~M$_{\oplus}$, and 3~M$_{\oplus}$. We also performed a set of simulations where we initially considered  three protoplanets rather than only two as in our nominal simulations. These simulations are discussed in Section~\ref{subsec:five+}. We stress that our choice of initial condition is ad-hoc and we do not model the early stages of planet formation where these initial objects grow in the disk. Our initial conditions were essentially motivated by the giant  impact scenarios explored in SPH simulations~\citep{kegerreisetal18,reinhardtetal20,rufucanup22}.

Our small embryos and also protoplanets are initially placed on orbits exterior to Saturn. The innermost body is placed between 5 to 10 Hill radii from Saturn and we arrange the subsequent bodies near different mean motion resonances with each other. Initial eccentricities of protoplanets and small embryos are randomly selected between  $10^{-3}$ and $10^{-2}$ and orbital inclinations between $10^{-5}$ and $10^{-2}$ degrees. Orbital element angles are randomly selected between 0 and 360 degrees.

\begin{table}
\centering
\begin{tabular}{@{}lcccc@{}}
\toprule
HMR sets & M$_{\textrm{small}}$ ~(M$_{\oplus}$) & N$_{\textrm{small}}$ & M$_{\textrm{p-Uranus}}$  & M$_{\textrm{p-Neptune}}$ \\ \midrule
0.5:5  & \multirow{3}{*}{0.5} & 5      & \multirow{3}{*}{13.5} & \multirow{3}{*}{16.5} \\
0.5:10 &                      & 10     &                       &                       \\
0.5:20 &                      & 20     &                       &                       \\ \midrule
1:5  & \multirow{3}{*}{1}   & 5      & \multirow{3}{*}{13}   & \multirow{3}{*}{16}   \\
1:10 &                      & 10     &                       &                       \\
1:20 &                      & 20     &                       &                       \\ \midrule
2:5  & \multirow{3}{*}{2}   & 5      & \multirow{3}{*}{12}   & \multirow{3}{*}{14}   \\
2:10 &                      & 10     &                       &                       \\
2:20 &                      & 20     &                       &                       \\ \midrule
3:5  & \multirow{3}{*}{3}   & 5      & \multirow{3}{*}{11}   & \multirow{3}{*}{13}   \\
3:10 &                      & 10     &                       &                       \\
3:20 &                      & 20     &                       &                       \\ 
\midrule
I15 sets & M$_{\textrm{embryo}}$ ~(M$_{\oplus}$) & N$_{\textrm{embryo}}$ &  &  \\ \midrule
6:10 & 6   & 10   &  -  &  -  \\
4-8:10 & 4 to 8   & 10   &  -  &  -  \\
\bottomrule
\end{tabular}
\caption{Table summarizing the \textcolor{black}{initial conditions of our simulations sets. The first 10 sets correspond to high mass ratio (HMR) scenario and the last 2 sets to I15 scenario. The columns are set name, initial mass of the embryos, initial number of embryos, proto-Uranus initial mass, and proto-Neptune initial mass.}}
\label{tab:sets}
\end{table}

Table \ref{tab:sets} shows all 12 sets of simulations from the high impact-to-target mass ratio scenario (hereafter, HMR scenario) considered in this paper. The columns are, from left-to-right, model name, initial mass of the embryos, initial number of embryos, proto-Uranus initial mass, and proto-Neptune initial mass. For each model, we have performed 1000 simulations. To compare the outcomes of the high mass ratio scenario with that of I15, we conducted two sets of simulations where planetary objects follow the initial distribution of I15. The first set considers 10 embryos of 6~M$_{\oplus}$ each, and the second considers 10 embryos with masses randomly selected between 4 and 8~M$_{\oplus}$ (see bottom of Table \ref{tab:sets}). In these complementary simulations of I15 scenario, Jupiter and Saturn start as in HMR simulations, at 5.2~au. Our simulations differ from those of I15's paper, where most simulations initiated with Jupiter at 3.5~au rather than at 5.2~au.

\section{Results}\label{sec:results}

We start the analysis of our results by tracking the dynamical evolution and collision history of all protoplanets and embryos in our simulations. Then, we classify each simulation within one of following five  categories as outlined below. 

\begin{itemize}
    \item \textit{Lost protoUN}: less than two protoplanets form (proto-Uranus or proto-Neptune are lost due to ejections or impacts with Jupiter/Saturn) at the end of the simulation.
    \item \textit{Jumpers}: an embryo (or protoplanet) is scattered and implanted into the inner Solar System or the final system is inconsistent with the Solar System (e.g. Saturn is ejected)
    \item \textit{At least one collide}: at least one protoplanet (proto-Uranus or proto-Neptune) collided with small embryos, and survived until the end of the simulation.
    \item \textit{Both collide}: proto-Uranus and proto-Neptune collided with at least one small embryo each and survived until the end of the simulation.
    \item \textit{No collision}: no collisions between protoplanets (proto-Uranus and proto-Neptune) and small embryos.
\end{itemize}

We are particularly interested in simulations in which both protoplanets (proto-Uranus and proto-Neptune) undergo at least one giant impact each and, obviously, both planets survive until the end of the simulation. This outcome corresponds to the case \textit{Both collide}, above. In our analysis, we are also interested in rejecting those cases where the broad orbital architecture of the Solar System was not preserved. This includes cases where either Jupiter or Saturn were lost due to instabilities, as well as those where small embryos or protoplanets are scattered and implanted into  the inner Solar System (see our text for \textit{Jumpers} definition). Simulations classified as \textit{Both collide}, \textit{At least one collide} or \textit{No collision} maintain both protoplanets, and also Jupiter and Saturn at the end of simulation and have no \textit{Jumpers}.

\noindent{
\begin{figure*}
    \centering
    \includegraphics[scale=0.85]{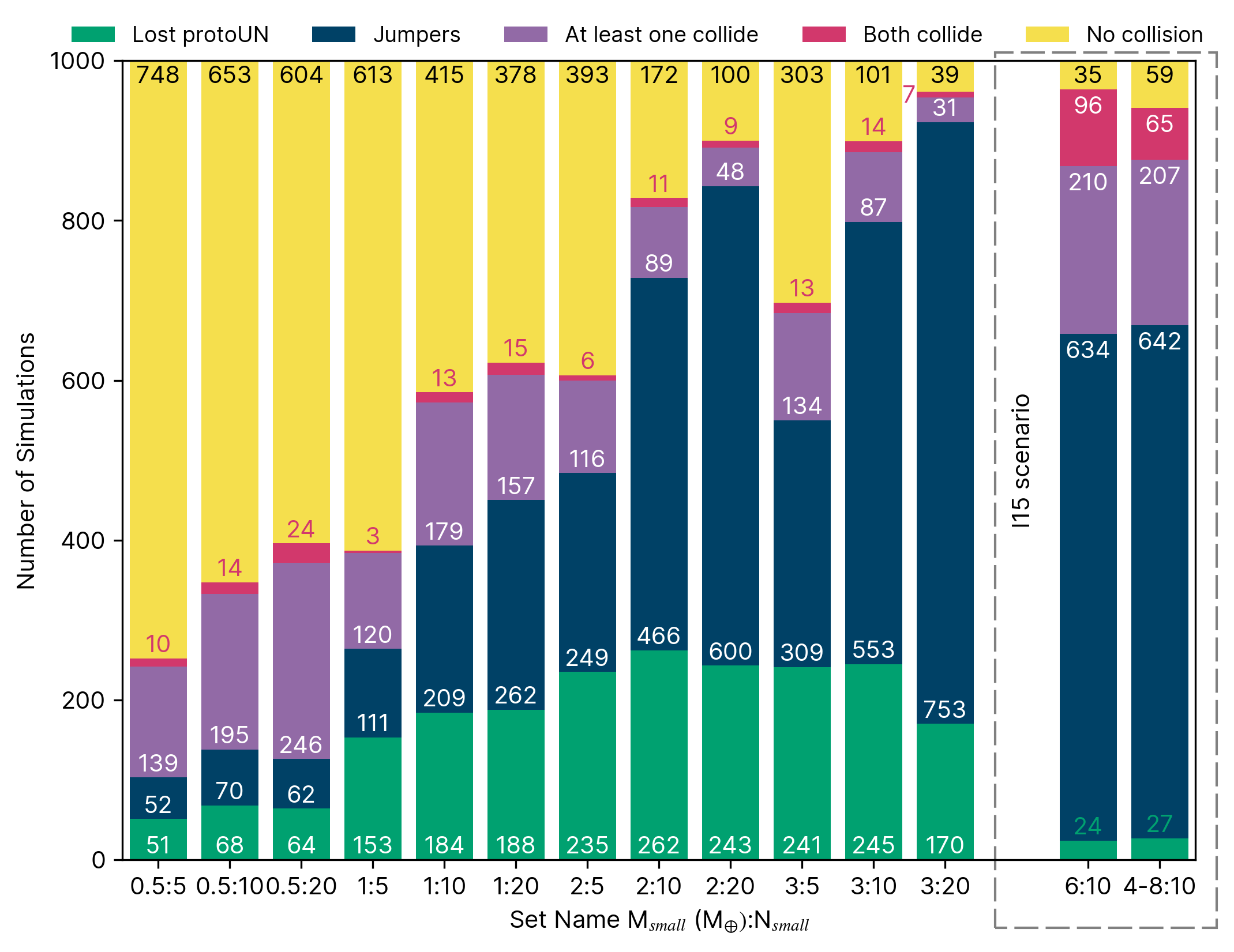}
    \caption{Stacked bars illustrating the results of 1000 numerical simulations for each set. The set name indicates the small embryos mass in M$_{\oplus}$ on the right and the number of small embryos on the left. The bars colour-coding denotes the final classification of simulations. The number indicates how many simulations in the set have the given category. The last two bars are the sets from I15 scenario.}
    \label{fig:stats}
\end{figure*}
}

Figure \ref{fig:stats} summaries the results of all our simulations and gives the number of systems within category of outcomes, as defined above. The x-axis shows the model name (see \ref{tab:sets}). Each vertical bar represents 1000 simulations. Colour-coded sub-bars shows the relative number of simulations within different categories of outcomes. 

In red, we show the number of simulations with the outcome \textit{Both Collide}.  The number of \textit{Both Collide} outcomes is somewhat similar among the different sets of our HMR scenario simulations, within a factor of 10, but in most cases within a factor of a few. The largest difference is a factor of 8 between the the set 0.5:20 (2.4\%), and the set, 1:5 (0.3\%). Overall, we found that about 1.16\% of our high mass ratio scenario simulations meet the \textit{Both Collide} criteria. This fraction is approximately ten times smaller than the 10.25\% obtained for the I15 scenario. Such difference indicates that collisions are more frequent in the I15 scenario, which involves larger embryos compared to the HMR scenario.

In fact, from the HMR scenario simulations, it is also clear that the number of collisions of small embryos with both protoplanets increases for simulations with larger M$_{\textrm{small}}$ (and also N$_{\textrm{small}}$). This trend is expected as larger embryos experience stronger migration and damping of orbital inclinations due to interactions with the gas. As a result, the orbits of more massive embryos tend to be more easily radially confined beyond Saturn (due to convergent migration) which increases the probabilities of giant impacts.

One could envision that increasing N$_{\textrm{small}}$ would also increase the probability of collisions, but this is not necessarily true because it may come with the expense of also increasing the level of dynamical scattering in the system and the potential number of \textit{Jumpers} (for a same given M$_{\textrm{small}}$). For instance, models where M$_{\textrm{small}}$ is set to 2 or 3~$M_{\oplus}$ show that the overall probability of collision decreases when one increases N$_{\textrm{small}}$. However, this same trend is not observed in the M$_{\textrm{small}}$ = 0.5~$M_{\oplus}$ set of simulations, where an initial larger number of embryos indeed resulted in more collisions (larger number of \textit{Both Collide} outcomes). This difference probably comes from the different levels of dynamical excitation that the system reach during their evolution. 

Figure \ref{fig:stats} also shows that the number of systems in the category \textit{Jumpers} tend to be higher in simulations with larger M$_{\textrm{small}}$. This occurs because larger M$_{\textrm{small}}$ embryos scattered into the inner solar system are more likely to have their orbits efficiently damped by the gas and decoupled from Jupiter~\citep{izidoroetal15} before ejections can take place. Finally, we also found that the number of systems with at least one collision between small embryos and protoplanets (purple and red bars combined in Figure \ref{fig:stats}) tend to slightly increase for simulations with smaller M$_{\textrm{small}}$.

\subsection{Simulations with Uranus and Neptune analogues} \label{sec:analogues}

\noindent{
\begin{figure*}
    \centering
    \includegraphics[scale=0.57]{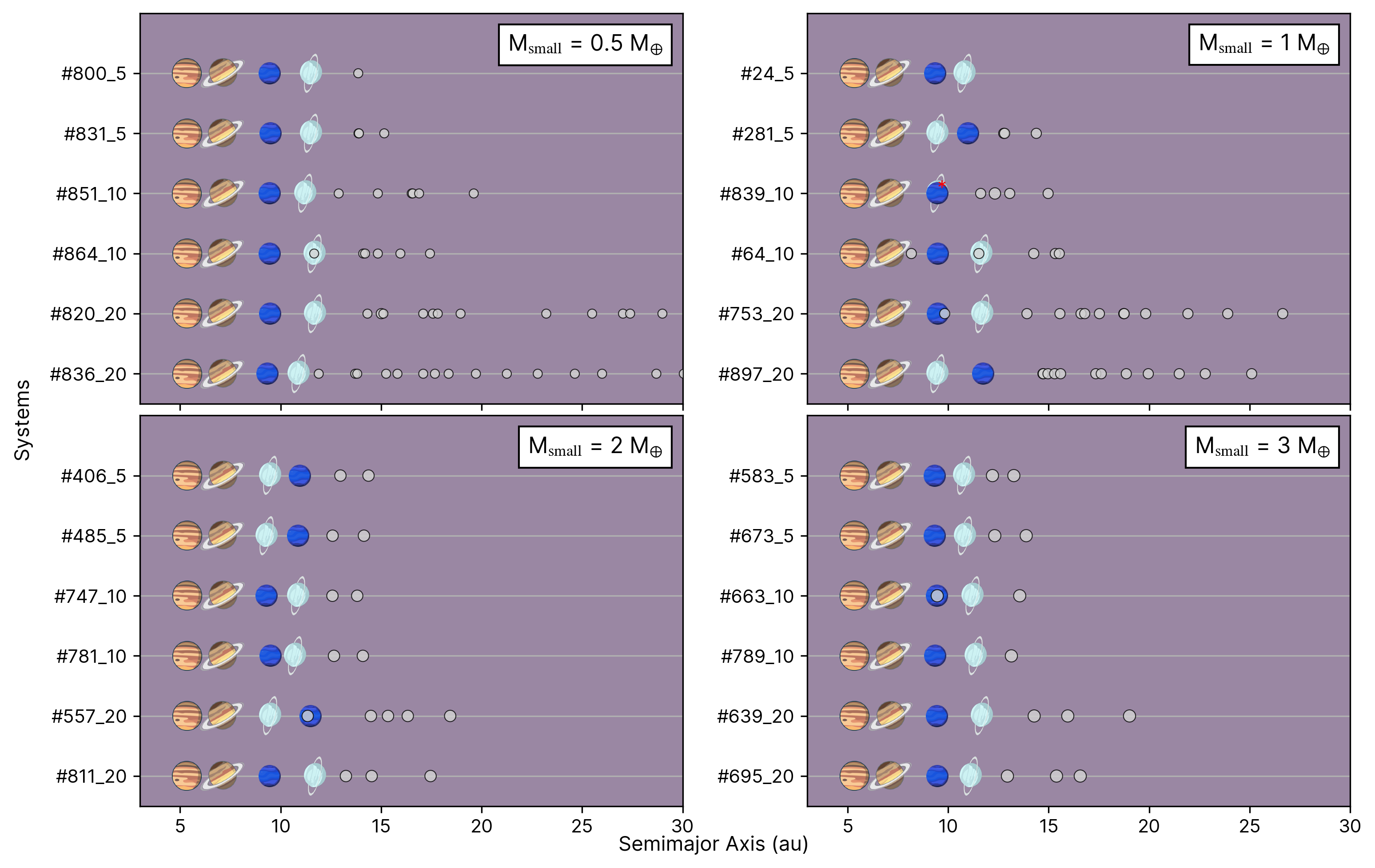}
    \caption{Final snapshot of simulations in which both protoplanets have experienced at least one collision. Each simulation is identified by $\#iii\_jj$, where $iii$ is the simulation number, and $jj$ is the number of small embryos that starts the simulation. The planets are sorted by their semi-major axis. The first planets are Jupiter and Saturn followed by proto-Uranus (light-blue) or proto-Neptune (dark-blue) analogues. The small embryos are represented by light-gray filled circles. The panels are arranged according to the M$_{\textrm{small}}$, with 0.5~M$_{\oplus}$, 1~M$_{\oplus}$, 2~M$_{\oplus}$, and 3~M$_{\oplus}$, respectively. * indicates co-orbital bodies.}
    \label{fig:sims}
\end{figure*}
}

In Figure \ref{fig:sims}, we plot the final dynamical architecture of 24 selected simulations where both proto-Uranus and proto-Neptune experienced at least one giant impact each (\textit{Both Collide} outcomes). We selected two simulations for each set of the HMR scenario (see Table \ref{tab:sets}). We group our simulations in different panels as indicated by the parameter M$_{\textrm{small}}$, shown on the top-left side of each panel. 

\textcolor{black}{It is important to keep in mind that the final time of our simulations represents the pre-instability configuration of the outer Solar System, characterized by more compact orbits for the giant planets compared to those of the present-day. It is well known that the four giant planets could have formed in a more compact configuration and later migrated to their current orbits due to gravitational interactions with a planetesimal disk beyond Neptune~\citep{fernandezip84,malhotra93,tsiganisetal05,nesvorny11}. Notably, a few systems in Figure \ref{fig:sims} show Uranus and Neptune in swapped orbits compare to their current configuration. During the instability phase, it is worth noting that Uranus and Neptune may have exchanged their orbital positions (e.g., Figure 4 in \citealt{nesvornymorbidelli12}).}

In the top panels of Figure \ref{fig:sims}, we observe that small embryos (in grey) often survive until the end of the simulations, with the number of leftover embryos increasing with N$_{\textrm{small}}$. This result is also supported by those of I15. As explained before, this is probably a consequence of limited radial migration and fewer ejection events in simulations with lower M$_{\textrm{small}}$. Simulations where N$_{\textrm{small}} \geq 10$ and M$_{\textrm{small}}\lesssim1~M_{\oplus}$ exhibit an average of about 3.6 ejections per simulation. Conversely, in sets with M$_{\textrm{small}}$ of 2~M$_{\oplus}$ and 3~M$_{\oplus}$ where N$_{\textrm{small}} \geq 10$, we observe an average of about 7.8 ejections per simulation. A large number of leftover embryos tend to survive in simulations with fewer ejections. It would be interesting to investigate in future studies whether scenarios with a large number of leftover embryos beyond Neptune are consistent with the dynamical evolution of the solar system~\citep{nesvornymorbidelli12,deiennoetal17} or not. In simulations with N$_{\textrm{small}}$ equal to 5 and 10, fewer small embryos tend to survive, typically less than 50\% of the initial number. Interestingly, some of our simulations also show co-orbital planets with Uranus and Neptune (see Figure \ref{fig:sims}).

In the following section, we analyse the rotation periods of the planets that experienced at least one collision. In fact, most planets experienced a single collision. Less than 5\% of the planets that suffered one collision experienced a second one in simulations with M$_{\textrm{small}}$ lower than 1~M$_{\oplus}$. In systems with M$_{\textrm{small}}$ higher than 1~M$_{\oplus}$ this ratio was slightly larger, about 15\%. Overall, secondary collisions are rare in our HMR simulations.

\subsection{Rotation Period}\label{subsec:rotation}

We calculate the rotation periods and obliquities of our final planets via post-processing of the impact history of our simulations. To do so, we store the positions, velocities, and masses of bodies involved in collisions at the moment of the impact. In order to precisely determine the impact geometry at the very moment when the bodies touch each other in a collision, we use the Bulirsch-Stoer integrator with accuracy parameter of $10^{-14}$. This is required because in some cases the numerical integrator identifies a collision when bodies have their physical radius already ``overlapping''. In this case, we integrate back in time the bodies involved in the collision until they reach the impact moment.

The rotation periods and obliquities of colliding planets are determined using two-body approximation. We have tested two different methods. In the first one we estimate the rotation period using the angular momentum conservation of a rigid sphere. In this case, the angular velocity of the resulting body is calculated as the ratio of the spin vector's magnitude by the moment of inertia of the sphere. The second method invokes a Maclaurin spheroid. In this case the angular velocity is given by the spin vector's magnitude divided by the Maclaurin spheroid's moment of inertia. We found that both methods produce very similar results, and we decide to use Maclaurin approximation. Although our method to estimate obliquity and rotation period is simplified, \citet{chauetal21} showed that this approximation have a good agreement with the results of SPH simulations in the case of our simulations. 

\noindent{
\begin{figure}
    \centering
    \includegraphics[scale=0.52]{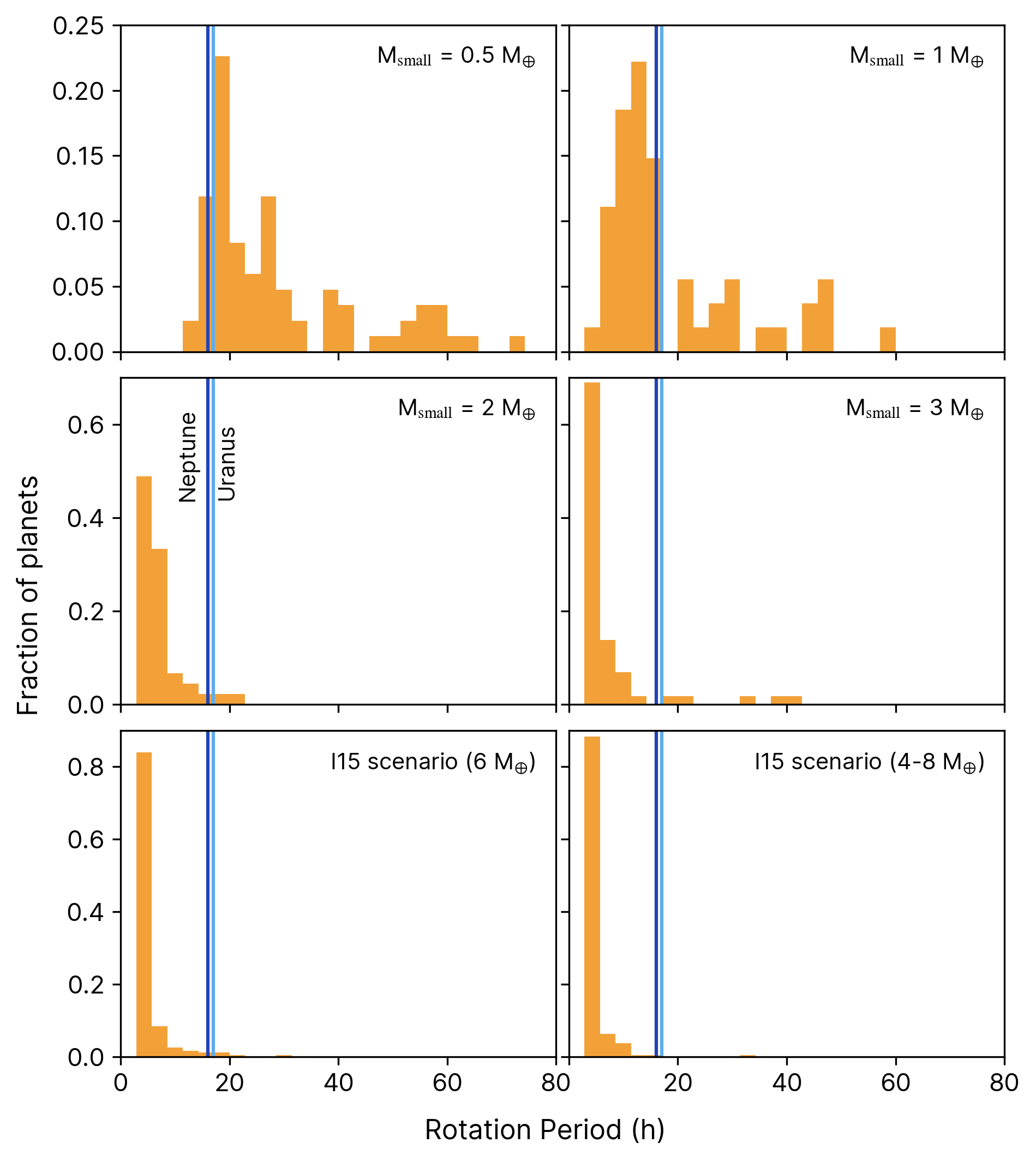}
    \caption{Fraction of planets analogues to Uranus/Neptune from \textit{Both Collide} simulations by their rotation period in hours. The light-blue and dark-blue reference lines represent the rotation periods of Uranus and Neptune, respectively. The four upper panels show planets that collided with specific small embryos with M$_{\textrm{small}}$ in HMR scenario simulations. The bottom panels show planets from I15 scenario with M$_{\textrm{embryo}}$ equal to 6~M$_{\oplus}$ and 4 to 8~M$_{\oplus}$.}
    \label{fig:rotation}
\end{figure}
}

Figure \ref{fig:rotation} shows the rotation periods distributions of final planets in our \textit{Both Collide} simulations. In the top-left panel of Figure \ref{fig:rotation}, we observe that the majority of planets formed by collisions where M$_{\textrm{small}}$ = 0.5~M$_{\oplus}$ and 1~M$_{\oplus}$ have rotation periods that fairly match the current values of Uranus and Neptune. As M$_{\textrm{small}}$ increases, final planets tend to rotate faster.  This is consistent with the results of SPH simulations~\citep{reinhardtetal20,chauetal21}. We recall the reader that we assume that at the beginning of our simulations protoplanets have no initial rotation. \textcolor{black}{ In order to test the effects of this assumption, we have also analysed a subset of our simulations considering that protoplanets may have some initial rotation from the process of accretion itself~\citep{visseretal20} (before giant impacts). We assumed a normal distribution for the initial spin rate, ranging from 70\% to 10\% of the breakup speed of colliding bodies and an isotropic distribution for the initial obliquity. Our results do not change qualitatively but the final distribution of rotation periods tend to be slightly shifted to shorter rotation periods.}

Table \ref{tab:rotmargin} shows the fraction of planets matching the rotation period of Uranus and Neptune (assuming an averaged value of 16.5~hours for both planets). We show the respective fraction of matches considering deviation intervals of $\pm15\%$, $\pm25\%$, $\pm50\%$ and $\pm100\%$, relative to the reference value. For the $\pm100\%$ interval range, we considered rotation periods between 5.7 hours (3 times the breakup speed of Uranus) and 33 hours (twice the averaged rotation period of Uranus and Neptune).

\begin{table}
\centering
\begin{tabular}{lcccc}
\hline
M$_{\textrm{small/embryo}}$ & $\pm 15\%$ (\%) & $\pm 25\%$ (\%) & $\pm 50\%$ (\%) & $\pm 100\%$ (\%) \\ \hline
0.5   & 26.19   & 38.1   & 48.81 &  70.24         \\
1    & 16.67   & 27.78   & 62.96  & 83.33         \\
2    & 2.17    & 6.52    & 19.57  & 50        \\
3    & 1.72    & 3.45   & 12.07  & 27.59        \\ \hline
6    & 1.69   & 3.8    & 7.17   & 16.03         \\
4 to 8  & 0.53  & 1.06  & 4.79   & 12.23       \\ \hline
\end{tabular}
\caption{Percentage of planets whose rotation periods fall within the margins of the mean values of Neptune and Uranus, approximately 16.5~h. \textcolor{black}{The columns show the mass of the colliding embryos, and percentage of simulations with objects where the  rotation periods is within $\pm 15\%$, $\pm 25\%$, $\pm 50\%$ and $\pm 100\%$, respectively.}}
\label{tab:rotmargin}
\end{table}

\textcolor{black}{One can observe in Table \ref{tab:rotmargin} a clear trend} -- the rotation periods of our final planets is shorter in scenarios where M$_{\textrm{small}}$ is larger. As mentioned before, this result is expected because of the transfer of linear momentum into angular momentum during collisions~\citep[e.g.][]{reinhardtetal20, chauetal21,rufucanup22}. This is also true for the simulations following the scenario of I15, where planets tend to rotate relatively even faster compared to those of the HMR scenario. 

Table \ref{tab:rotmargin} shows that up to 26\% (49\%) of the giant impacts in simulations assuming M$_{\textrm{small}}$ = 0.5~M$_{\oplus}$ produced final planets with rotation periods within 15\% (50\%) of that of Uranus and Neptune (averaged value). For the scenario simulated by I15 (M$_{\textrm{small}}$ = 6~M$_{\oplus}$) the corresponding fractions are 1.7\% and 7.17\%. 





\subsection{Obliquity}

In this section we calculate the final obliquities of our planets by measuring the angle of the spin vector relative to the planet's orbital angular momentum. We assume that protoplanets and embryos have \textcolor{black}{a negligible} initial rotation and we also assume that their initial spins are aligned with their orbital angular momentum. 

The obliquities of our final planets vary from 0 to 180 degrees. An obliquity between 0 and 90 degrees represents planets with prograde rotation, while an obliquity between 90 and 180 degrees represents planets with retrograde rotation.

\noindent{
\begin{figure}
    \centering
    \includegraphics[scale=0.62]{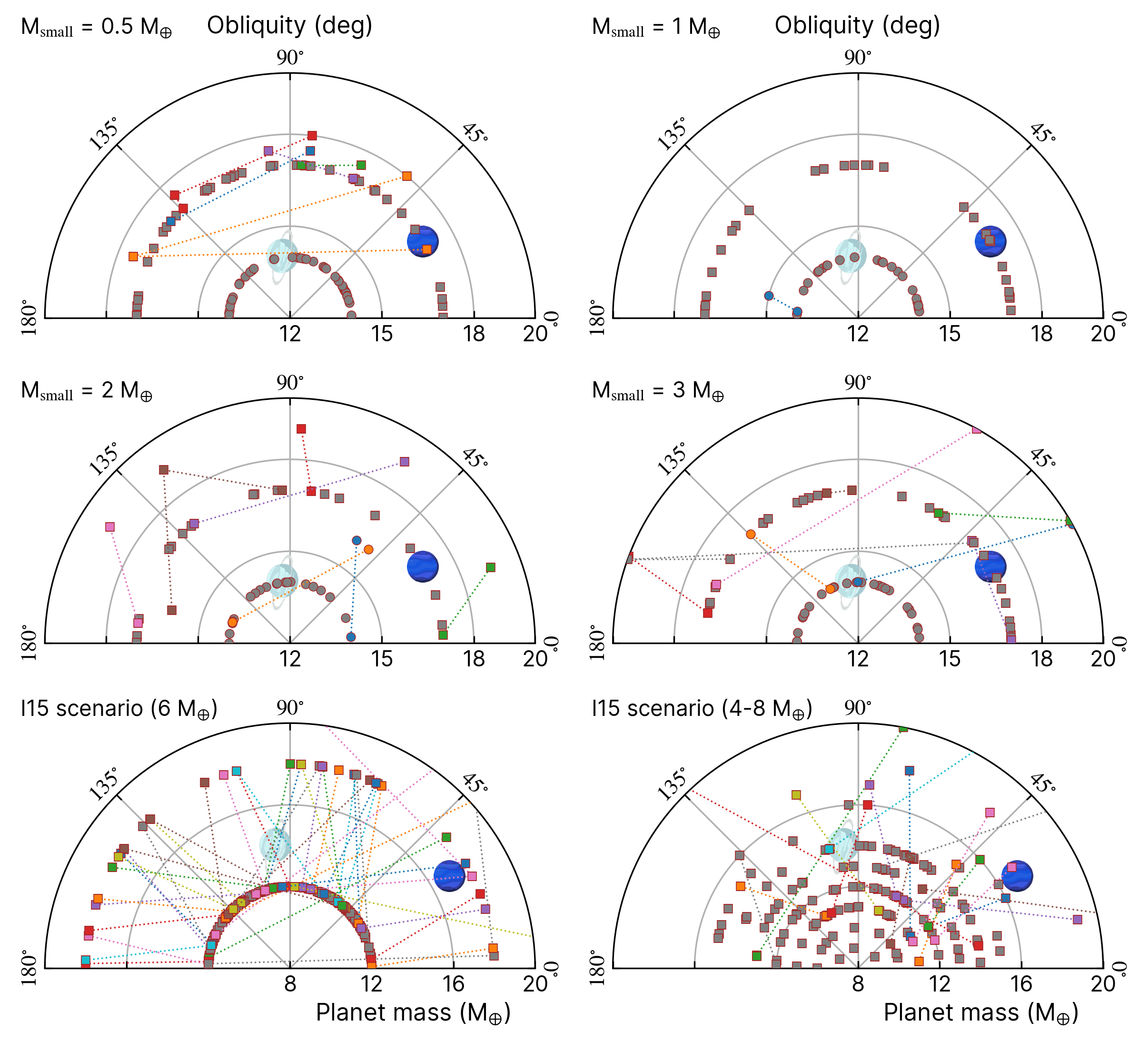}
    \caption{Obliquity and mass evolution of planets after collisions in HMR \textcolor{black}{and I15} scenarios. Each marker represent the resulting planet after a collision. Square markers represent Neptune analogues and circle markers Uranus analogues. In the radial axis we plot the planet mass. The angular axis shows the obliquity (e.g. for 0 deg in obliquity, the planet is aligned with system's plane). The panels indicate the different values of M$_{\textrm{small}}$. The grey markers denote planets that experienced only one collision in simulations. Coloured markers connected with dashed lines show the resulting masses and obliquities after planets' successive collisions. Uranus (light-blue) and Neptune (dark-blue) are plotted for reference.}
    \label{fig:obliquity}
\end{figure}
}

Figure \ref{fig:obliquity} shows the distribution of mass and obliquity our final planets. Each data point represents a collision, with square and circle markers denoting Neptune- and Uranus-mass-like planets, respectively. \textcolor{black}{The isolated gray points indicate planets that experienced a single collision, while the points connected by dashed lines correspond to planets that underwent multiple collisions.} The final obliquity distribution of the planets exhibits an almost isotropic distribution, as expected during the accretion phase involving collisions between embryos~\citep{kokuboetal07}. 


Figure \ref{fig:obliquity} also shows that that the majority of planets in our simulations experienced a single giant collision, as discussed in section \ref{sec:analogues}. Additionally, in very rare cases (orange dot in the top panel of Figure \ref{fig:obliquity}), some protoplanets experienced three collisions.

\section{Statistics} \label{sec:prob}

In this section we evaluate which scenario (HMR vs. I15) is statistically more likely to account for the formation of Uranus and Neptune within our planet formation model. We now  calculate the fraction of systems where 

\begin{itemize}
    \item  both proto-Uranus and proto-Neptune experience at least 1 collision with small embryos. For simulations in the I15 scenario, we impose that the Uranus and Neptune analogue  have masses larger than 12 M$_{\oplus}$ and experienced at least 1 collision with other embryos. 
    \item the rotation periods of the final planets are within $\pm15\%$, $\pm 25\%$, $\pm 50\%$ and $\pm 100\%$ of the averaged rotation period of Uranus and Neptune.
\end{itemize}

\begin{table}
\centering
\begin{tabular}{lcccc}
\hline
M$_{\textrm{small-embryo}}$ & $\pm 15\%$ (\%) & $\pm 25\%$ (\%) & $\pm 50\%$ (\%) & $\pm 100\%$ (\%) \\ \hline
0.5   & 0.42   & 0.61   & 0.78 &  1.12         \\
1    & 0.17   & 0.28   & 0.63  & 0.83        \\
2    & 0.02    & 0.06    & 0.18  & 0.45        \\
3    & 0.02    & 0.04   & 0.14  & 0.31      \\ \hline
6    & 0.16   & 0.36    & 0.69   & 1.54         \\
4 to 8  & 0.03  & 0.07  & 0.31   & 0.8       \\ \hline
\end{tabular}
\caption{Percentage of simulations broadly matching the pre-dynamical instability architecture of the outer solar system while ensuring that the resulting planets exhibit rotation periods consistent with those of the ice giants. \textcolor{black}{The columns show the mass of the colliding embryos, and percentage of simulations with objects where the rotation periods is within $\pm 15\%$, $\pm 25\%$, $\pm 50\%$ and $\pm 100\%$, respectively.}}
\label{tab:combprob}

\end{table}

Table \ref{tab:combprob} shows the fraction of systems matching these constraints simultaneously. As one can see, the success rates are generally low. For cases where the rotation period falls within $\pm 15\%$ of those of Uranus and Neptune only 0.1-0.5\% of the simulations matched these constraints. For cases where the rotation period falls within $\pm 50\%$ of those of Uranus and Neptune the success rate is of the order of 1\%. Overall, the simulations of HMR and I15 scenarios produce fairly similar results, with overall success rates within a factor or 2. These results suggest that from a planet formation perspective there is no clear preference of one of these scenarios over the other to account for the formation of Uranus and Neptune.

\subsection{Five or more giant planets scenario}\label{subsec:five+}

Our nominal simulations of the HMR scenario start with only two protoplanets (see Table \ref{tab:sets} for details). In this section we performed an additional set of simulations assuming a larger initial number of protoplanets (3 and 4). As in our nominal HMR simulations, the initial masses of the protoplanets are chosen such that a single giant impact is required to match the masses of present-day Uranus and Neptune. Our new set of simulations start with 3 or 4 protoplanets interacting with 5, 10 or 20 small embryos within the disk. This scenario is motivated by solar system evolution models suggesting that the solar system may have had five giant planets ~\citep[][]{nesvornymorbidelli12}.

The fraction of successful simulations matching the masses and rotation periods of Uranus and Neptune are very similar to those of our nominal HMR scenario (Table~\ref{tab:combprob}). However, we also noticed that in these simulations collisions among the protoplanets themselves are more common. Therefore, by increasing the number of additional protoplanets in our model does not seem to lead to any clear advantage. 

\section{Caveats}\label{sec:discussion}


\textcolor{black}{Our simulations are simplified in many aspects. For instance, we use a 1-D underlying disk to represent the protoplanetary disk. The simulations also start with fully formed Jupiter and Saturn (see also \cite{izidoroetal15}), and considering equal mass planetary objects beyond Jupiter and Saturn (in some cases). In future studies, it will be interesting to model the growth and migration of the gas giants simultaneously with the accretion of Uranus and Neptune and to avail different disk profiles.}

\textcolor{black}{As discussed before, our collision treatment in simulations neglects the effects of fragmentation and hit-and-run events. In order to assess the validity of this approximation, we calculated the impact velocity relative to the escape velocity for collisions in our simulations. We found that impact velocities are consistently lower than 1.05 \(v_{\text{esc}}\), ranging from 0.95 to 1.05 depending on \(M_{\text{small}}\). Smaller values of \(M_{\text{small}}\) (and larger \(N_{\text{small}}\)) tends to result in slightly higher impact velocities, as the orbital damping of eccentricity and inclination for smaller masses is relatively less efficient, allowing larger eccentricities, inclinations and relative velocities. Additionally, we verified that approximately 30\% of the collisions in our simulations occur with impact parameters greater than 0.8. Previous studies have shown that high impact parameters can lead to modest fragmentation ($\sim$10-20\% of the total colliding mass) even if impact velocities are relatively low, on the order of \(v_{\text{esc}}\) \cite[e.g.][]{reinhardtetal20}. This outcome can allow some angular momentum loss when fragments escape the colliding bodies, potentially increasing the rotation period of planets by up to a factor of $\sim$2 \cite[e.g.][]{reinhardtetal20}, on average. To account for this potential effect in a simplified and practical manner, we recalculated the probabilities of Table \ref{tab:combprob}, considering that grazing collisions with impact parameter larger than 0.8 could result in rotation periods a factor of 2 larger than that assumed for perfect merging events \citep{reinhardtetal20}. Our results show that under this assumption the probabilities of Table \ref{tab:combprob} changes by up to a factor of a few. However, our qualitative conclusions remains broadly unchanged.}

\textcolor{black}{Although this work does not address constraints on Uranus' proto-satellite disk and formation of its satellites, two competing scenarios are worth discussing here. The first one proposes that the proto-satellite disk resulted from a single impact with masses of approximately $2$–$3~M_{\oplus}$~\citep{reinhardtetal20, wooetal2022}. However, reproducing the compositions of Uranus' moons requires impactors composed entirely of rocky material~\citep{wooetal2022}, which contradicts the ice-dominated composition observed in bodies from this region. The second view combines the formation of a primordial satellite disk followed by a giant impact~\citep{morbidellietal12}. In this scenario, the impact leads to the destruction of primordial satellites and the formation of an outer disk. This disk is envisioned to contain the mass and composition necessary to form Uranus' moons but would be misaligned with the planet's new rotational axis. Moreover, aligning the satellites with Uranus' equator would require an secondary disk -- generated by the impact -- that is too massive to be explained by typical impactors (ranging from $0.5$ to $3~M_{\oplus}$) ~\citep{rufucanup22}. The origin of Uranus' moons remains an intriguing unresolved question, as neither large nor small impactors have provided a strong explanation thus far.}

\section{Summary}\label{sec:summary}

In this work we modelled the final accretion phase of Uranus and Neptune in a gaseous protoplanetary disk within the giant impact hypothesis using N-body numerical simulations. We investigated how different giant impacts scenarios affects the final rotation periods of Uranus and Neptune and their obliquities. Our simulations started with Jupiter and Saturn fully formed and a population of planetary objects beyond Saturn. In our simulations we envision that Uranus' and Neptune's precursors formed by pebble and/or planetesimals accretion and experienced a final phase of giant impacts. Our simulations test two competing giant impact scenarios to account for their origins. In the first one, we assumed that Uranus and Neptune precursors  formed from collisions with small embryos only, with masses between 0.5 and 3~M$_{\oplus}$. This scenario is refereed to as the high-mass-ratio scenario and is motivated by studies invoking SPH impact simulations \citep{reinhardtetal20,wooetal2022,rufucanup22}. In the second one, we assumed that Uranus and Neptune accreted from giant impacts among 5-10 protoplanets of about $\sim$6~M$_{\oplus}$ (or 4-8 $_{\oplus}$) each. In this model Uranus and Neptune accreted from collisions among objects with mass ratios of the order of unity \citep{izidoroetal15}.

While the scenario of I15 has shown to reproduce fairly well the masses of Uranus and Neptune, final planets produced in these simulations tend to have rotation periods too short compared to Uranus and Neptune \citep{chauetal21} due to conversion of linear momentum into angular momentum during the impact. Recent SPH simulations, on the other hand, show that Uranus and Neptune rotation periods are better reproduced in simulations with high target-to-impactor mass ratios (e.g. a proto-Uranus of 13~M$_{\oplus}$ and another object of 1~M$_{\oplus}$). Where this scenario is appealing, as demonstrated by SPH simulations, it has not been tested within the context of planet formation and dynamical evolution. In this work, we compared the relative successes rates of these competing scenarios using an accretion model for Uranus and Neptune.

Our simulations start with fully formed Jupiter and Saturn embedded in a gaseous protoplanetary disk. Beyond Saturn we consider a population or protoplanetary objects, the precursors of Uranus and Neptune.  We conducted 12 sets of simulations considering different combination of impactor masses and initial number of embryos (M$_{\textrm{small}}$ = 0.5, 1, 2, and 3~M$_{\oplus}$) and numbers of small embryos (N$_{\textrm{small}}$ = 5, 10, and 20). For each set of simulations we ran 1000 slightly different simulations by slightly changing the initial distribution of the embryos. 

As expected, the results of our simulations confirmed that the accretion of Uranus and Neptune from objects with large mass ratio leads to rotation periods more consistent with those of the real planets. However, we found that the probability of occurring collisions in these specific simulations is significantly reduced -- compared to scenarios allowing collision between objects with order of unity mass ratio. Gas tidal damping is less efficiently for embryos with masses lower than 1-3~M$_{\oplus}$ at the current location of Uranus and Neptune \textcolor{black}{reducing the system's confinement}. Consequently, dynamical scattering dominates over accretion and instead of colliding with the proto-Uranus and Neptune small mass objects are more likely to be dynamically scattered by them. 

Our results suggest that the probability of broadly reproducing the masses, mass ratio, and rotation period of Uranus and Neptune in both the high-mass-ratio's and in the I15's scenario is broadly similar, within a factor of 2, with probabilities of the order of 0.1 to 1\%. Altogether, these probabilities seem to suggest that the formation of Uranus and Neptune is a fortuitous outcome of the chaotic phase of planetary accretion. However, if there is a still unknown viable mechanism to slow down Uranus and Neptune after their accretion phase (e.g. interactions with the surrounding gas disk) the success rate of models where impactors have roughly equal (and large, of about 5~M$_{\oplus}$) masses becomes up to one order of magnitude higher.

\section*{Acknowledgements}
L.~E. acknowledges the financial support from FAPESP through grants 2021/00628-6 and 	
2023/09307-3.
A.~I. is thankful to the NASA Emerging Worlds Program for financial support (Grant n. 80NSSC23K0868).
O.~C.~W. thanks Conselho Nacional de Desenvolvimento Científico e Tecnológico (CNPq) proc. 305210/2018-1 and FAPESP proc. 2016/24561-0 for financial support.

\section*{Data availability}
The data underlying this article will be shared on reasonable request to the corresponding author.







\printcredits

\bibliographystyle{cas-model2-names}

\bibliography{references_updated}

\begin{thebibliography}{76}
\expandafter\ifx\csname natexlab\endcsname\relax\def\natexlab#1{#1}\fi
\providecommand{\url}[1]{\texttt{#1}}
\providecommand{\href}[2]{#2}
\providecommand{\path}[1]{#1}
\providecommand{\DOIprefix}{doi:}
\providecommand{\ArXivprefix}{arXiv:}
\providecommand{\URLprefix}{URL: }
\providecommand{\Pubmedprefix}{pmid:}
\providecommand{\doi}[1]{\href{http://dx.doi.org/#1}{\path{#1}}}
\providecommand{\Pubmed}[1]{\href{pmid:#1}{\path{#1}}}
\providecommand{\bibinfo}[2]{#2}
\ifx\xfnm\relax \def\xfnm[#1]{\unskip,\space#1}\fi
\bibitem[{{Adachi} et~al.(1976){Adachi}, {Hayashi} and {Nakazawa}}]{adachietal76}
\bibinfo{author}{{Adachi}, I.}, \bibinfo{author}{{Hayashi}, C.}, \bibinfo{author}{{Nakazawa}, K.}, \bibinfo{year}{1976}.
\newblock \bibinfo{title}{{The gas drag effect on the elliptical motion of a solid body in the primordial solar nebula.}}
\newblock \bibinfo{journal}{Progress of Theoretical Physics} \bibinfo{volume}{56}, \bibinfo{pages}{1756--1771}.
\newblock \DOIprefix\doi{10.1143/PTP.56.1756}.
\bibitem[{{Biersteker} and {Schlichting}(2019)}]{bierstekerschlichting19}
\bibinfo{author}{{Biersteker}, J.B.}, \bibinfo{author}{{Schlichting}, H.E.}, \bibinfo{year}{2019}.
\newblock \bibinfo{title}{{Atmospheric mass-loss due to giant impacts: the importance of the thermal component for hydrogen-helium envelopes}}.
\newblock \bibinfo{journal}{Monthly Notices of the Royal Astronomical Society} \bibinfo{volume}{485}, \bibinfo{pages}{4454--4463}.
\newblock \DOIprefix\doi{10.1093/mnras/stz738}, \href{http://arxiv.org/abs/1809.06810}{\tt arXiv:1809.06810}.
\bibitem[{{Bitsch} et~al.(2015){Bitsch}, {Lambrechts} and {Johansen}}]{bitschetal15b}
\bibinfo{author}{{Bitsch}, B.}, \bibinfo{author}{{Lambrechts}, M.}, \bibinfo{author}{{Johansen}, A.}, \bibinfo{year}{2015}.
\newblock \bibinfo{title}{{The growth of planets by pebble accretion in evolving protoplanetary discs}}.
\newblock \bibinfo{journal}{Astronomy \& Astrophysics} \bibinfo{volume}{582}, \bibinfo{pages}{A112}.
\newblock \DOIprefix\doi{10.1051/0004-6361/201526463}, \href{http://arxiv.org/abs/1507.05209}{\tt arXiv:1507.05209}.
\bibitem[{{Bitsch} et~al.(2018){Bitsch}, {Morbidelli}, {Johansen}, {Lega}, {Lambrechts} and {Crida}}]{bitschetal18}
\bibinfo{author}{{Bitsch}, B.}, \bibinfo{author}{{Morbidelli}, A.}, \bibinfo{author}{{Johansen}, A.}, \bibinfo{author}{{Lega}, E.}, \bibinfo{author}{{Lambrechts}, M.}, \bibinfo{author}{{Crida}, A.}, \bibinfo{year}{2018}.
\newblock \bibinfo{title}{{Pebble-isolation mass: Scaling law and implications for the formation of super-Earths and gas giants}}.
\newblock \bibinfo{journal}{Astronomy \& Astrophysics} \bibinfo{volume}{612}, \bibinfo{pages}{A30}.
\newblock \DOIprefix\doi{10.1051/0004-6361/201731931}, \href{http://arxiv.org/abs/1801.02341}{\tt arXiv:1801.02341}.
\bibitem[{{Bou{\'e}} and {Laskar}(2010)}]{bouelaskar10}
\bibinfo{author}{{Bou{\'e}}, G.}, \bibinfo{author}{{Laskar}, J.}, \bibinfo{year}{2010}.
\newblock \bibinfo{title}{{A Collisionless Scenario for Uranus Tilting}}.
\newblock \bibinfo{journal}{The Astrophysical Journal Letters} \bibinfo{volume}{712}, \bibinfo{pages}{L44--L47}.
\newblock \DOIprefix\doi{10.1088/2041-8205/712/1/L44}, \href{http://arxiv.org/abs/0912.0181}{\tt arXiv:0912.0181}.
\bibitem[{Chambers(1999)}]{chambers99}
\bibinfo{author}{Chambers, J.E.}, \bibinfo{year}{1999}.
\newblock \bibinfo{title}{A hybrid symplectic integrator that permits close encounters between massive bodies}.
\newblock \bibinfo{journal}{Monthly Notices of the Royal Astronomical Society} \bibinfo{volume}{304}, \bibinfo{pages}{793--799}.
\bibitem[{{Chambers}(2001)}]{chambers01}
\bibinfo{author}{{Chambers}, J.E.}, \bibinfo{year}{2001}.
\newblock \bibinfo{title}{{Making More Terrestrial Planets}}.
\newblock \bibinfo{journal}{Icarus} \bibinfo{volume}{152}, \bibinfo{pages}{205--224}.
\newblock \DOIprefix\doi{10.1006/icar.2001.6639}.
\bibitem[{{Chau} et~al.(2021){Chau}, {Reinhardt}, {Izidoro}, {Stadel} and {Helled}}]{chauetal21}
\bibinfo{author}{{Chau}, A.}, \bibinfo{author}{{Reinhardt}, C.}, \bibinfo{author}{{Izidoro}, A.}, \bibinfo{author}{{Stadel}, J.}, \bibinfo{author}{{Helled}, R.}, \bibinfo{year}{2021}.
\newblock \bibinfo{title}{{Could Uranus and Neptune form by collisions of planetary embryos?}}
\newblock \bibinfo{journal}{Monthly Notices of the Royal Astronomical Society} \bibinfo{volume}{502}, \bibinfo{pages}{1647--1660}.
\newblock \DOIprefix\doi{10.1093/mnras/staa4021}, \href{http://arxiv.org/abs/2009.10100}{\tt arXiv:2009.10100}.
\bibitem[{{Cresswell} and {Nelson}(2006)}]{cresswellnelson06}
\bibinfo{author}{{Cresswell}, P.}, \bibinfo{author}{{Nelson}, R.P.}, \bibinfo{year}{2006}.
\newblock \bibinfo{title}{{On the evolution of multiple protoplanets embedded in a protostellar disc}}.
\newblock \bibinfo{journal}{Astronomy \& Astrophysics} \bibinfo{volume}{450}, \bibinfo{pages}{833--853}.
\newblock \DOIprefix\doi{10.1051/0004-6361:20054551}.
\bibitem[{{Cresswell} and {Nelson}(2008)}]{cresswellnelson08}
\bibinfo{author}{{Cresswell}, P.}, \bibinfo{author}{{Nelson}, R.P.}, \bibinfo{year}{2008}.
\newblock \bibinfo{title}{{Three-dimensional simulations of multiple protoplanets embedded in a protostellar disc}}.
\newblock \bibinfo{journal}{Astronomy \& Astrophysics} \bibinfo{volume}{482}, \bibinfo{pages}{677--690}.
\newblock \DOIprefix\doi{10.1051/0004-6361:20079178}, \href{http://arxiv.org/abs/0811.4322}{\tt arXiv:0811.4322}.
\bibitem[{{Crida} et~al.(2008){Crida}, {S{\'a}ndor} and {Kley}}]{cridaetal08}
\bibinfo{author}{{Crida}, A.}, \bibinfo{author}{{S{\'a}ndor}, Z.}, \bibinfo{author}{{Kley}, W.}, \bibinfo{year}{2008}.
\newblock \bibinfo{title}{{Influence of an inner disc on the orbital evolution of massive planets migrating in resonance}}.
\newblock \bibinfo{journal}{A\&A} \bibinfo{volume}{483}, \bibinfo{pages}{325--337}.
\newblock \DOIprefix\doi{10.1051/0004-6361:20079291}, \href{http://arxiv.org/abs/0802.2014}{\tt arXiv:0802.2014}.
\bibitem[{{{\'C}uk} and {Stewart}(2012)}]{cukstewart12}
\bibinfo{author}{{{\'C}uk}, M.}, \bibinfo{author}{{Stewart}, S.T.}, \bibinfo{year}{2012}.
\newblock \bibinfo{title}{{Making the Moon from a Fast-Spinning Earth: A Giant Impact Followed by Resonant Despinning}}.
\newblock \bibinfo{journal}{Science} \bibinfo{volume}{338}, \bibinfo{pages}{1047}.
\newblock \DOIprefix\doi{10.1126/science.1225542}.
\bibitem[{{Deienno} et~al.(2017){Deienno}, {Morbidelli}, {Gomes} and {Nesvorn{\'y}}}]{deiennoetal17}
\bibinfo{author}{{Deienno}, R.}, \bibinfo{author}{{Morbidelli}, A.}, \bibinfo{author}{{Gomes}, R.S.}, \bibinfo{author}{{Nesvorn{\'y}}, D.}, \bibinfo{year}{2017}.
\newblock \bibinfo{title}{{Constraining the Giant Planets{\textquoteright} Initial Configuration from Their Evolution: Implications for the Timing of the Planetary Instability}}.
\newblock \bibinfo{journal}{The Astronomical Journal} \bibinfo{volume}{153}, \bibinfo{pages}{153}.
\newblock \DOIprefix\doi{10.3847/1538-3881/aa5eaa}, \href{http://arxiv.org/abs/1702.02094}{\tt arXiv:1702.02094}.
\bibitem[{{Dodson-Robinson} and {Bodenheimer}(2010)}]{dodsonrobinsonbodenheimer10}
\bibinfo{author}{{Dodson-Robinson}, S.E.}, \bibinfo{author}{{Bodenheimer}, P.}, \bibinfo{year}{2010}.
\newblock \bibinfo{title}{{The formation of Uranus and Neptune in solid-rich feeding zones: Connecting chemistry and dynamics}}.
\newblock \bibinfo{journal}{Icarus} \bibinfo{volume}{207}, \bibinfo{pages}{491--498}.
\newblock \DOIprefix\doi{10.1016/j.icarus.2009.11.021}, \href{http://arxiv.org/abs/0911.3873}{\tt arXiv:0911.3873}.
\bibitem[{{Dr{\k{a}}{\.z}kowska} et~al.(2016){Dr{\k{a}}{\.z}kowska}, {Alibert} and {Moore}}]{drazkowskaetal16}
\bibinfo{author}{{Dr{\k{a}}{\.z}kowska}, J.}, \bibinfo{author}{{Alibert}, Y.}, \bibinfo{author}{{Moore}, B.}, \bibinfo{year}{2016}.
\newblock \bibinfo{title}{{Close-in planetesimal formation by pile-up of drifting pebbles}}.
\newblock \bibinfo{journal}{Astronomy \& Astrophysics} \bibinfo{volume}{594}, \bibinfo{pages}{A105}.
\newblock \DOIprefix\doi{10.1051/0004-6361/201628983}, \href{http://arxiv.org/abs/1607.05734}{\tt arXiv:1607.05734}.
\bibitem[{{Dr{\k{a}}{\.z}kowska} and {Dullemond}(2018)}]{drazkowskadullemond18}
\bibinfo{author}{{Dr{\k{a}}{\.z}kowska}, J.}, \bibinfo{author}{{Dullemond}, C.P.}, \bibinfo{year}{2018}.
\newblock \bibinfo{title}{{Planetesimal formation during protoplanetary disk buildup}}.
\newblock \bibinfo{journal}{Astronomy \& Astrophysics} \bibinfo{volume}{614}, \bibinfo{pages}{A62}.
\newblock \DOIprefix\doi{10.1051/0004-6361/201732221}, \href{http://arxiv.org/abs/1803.00575}{\tt arXiv:1803.00575}.
\bibitem[{{Fernandez} and {Ip}(1984)}]{fernandezip84}
\bibinfo{author}{{Fernandez}, J.A.}, \bibinfo{author}{{Ip}, W.H.}, \bibinfo{year}{1984}.
\newblock \bibinfo{title}{{Some dynamical aspects of the accretion of Uranus and Neptune: The exchange of orbital angular momentum with planetesimals}}.
\newblock \bibinfo{journal}{Icarus} \bibinfo{volume}{58}, \bibinfo{pages}{109--120}.
\newblock \DOIprefix\doi{10.1016/0019-1035(84)90101-5}.
\bibitem[{{Goldreich} and {Tremaine}(1980)}]{goldreichtremaine80}
\bibinfo{author}{{Goldreich}, P.}, \bibinfo{author}{{Tremaine}, S.}, \bibinfo{year}{1980}.
\newblock \bibinfo{title}{{Disk-satellite interactions.}}
\newblock \bibinfo{journal}{The Astrophysical Journal} \bibinfo{volume}{241}, \bibinfo{pages}{425--441}.
\newblock \DOIprefix\doi{10.1086/158356}.
\bibitem[{{Guilera} and {S{\'a}ndor}(2017)}]{guilerasandor17}
\bibinfo{author}{{Guilera}, O.M.}, \bibinfo{author}{{S{\'a}ndor}, Z.}, \bibinfo{year}{2017}.
\newblock \bibinfo{title}{{Giant planet formation at the pressure maxima of protoplanetary disks}}.
\newblock \bibinfo{journal}{Astronomy \& Astrophysics} \bibinfo{volume}{604}, \bibinfo{pages}{A10}.
\newblock \DOIprefix\doi{10.1051/0004-6361/201629843}, \href{http://arxiv.org/abs/1610.01232}{\tt arXiv:1610.01232}.
\bibitem[{{Helled}(2023)}]{helled23}
\bibinfo{author}{{Helled}, R.}, \bibinfo{year}{2023}.
\newblock \bibinfo{title}{{The mass of gas giant planets: Is Saturn a failed gas giant?}}
\newblock \bibinfo{journal}{Astronomy \& Astrophysics} \bibinfo{volume}{675}, \bibinfo{pages}{L8}.
\newblock \DOIprefix\doi{10.1051/0004-6361/202346850}, \href{http://arxiv.org/abs/2306.14740}{\tt arXiv:2306.14740}.
\bibitem[{{Helled} et~al.(2011){Helled}, {Anderson}, {Podolak} and {Schubert}}]{helledetal11}
\bibinfo{author}{{Helled}, R.}, \bibinfo{author}{{Anderson}, J.D.}, \bibinfo{author}{{Podolak}, M.}, \bibinfo{author}{{Schubert}, G.}, \bibinfo{year}{2011}.
\newblock \bibinfo{title}{{Interior Models of Uranus and Neptune}}.
\newblock \bibinfo{journal}{The Astrophysical Journal} \bibinfo{volume}{726}, \bibinfo{pages}{15}.
\newblock \DOIprefix\doi{10.1088/0004-637X/726/1/15}, \href{http://arxiv.org/abs/1010.5546}{\tt arXiv:1010.5546}.
\bibitem[{{Helled} and {Bodenheimer}(2014)}]{helledbodenheimer14}
\bibinfo{author}{{Helled}, R.}, \bibinfo{author}{{Bodenheimer}, P.}, \bibinfo{year}{2014}.
\newblock \bibinfo{title}{{The Formation of Uranus and Neptune: Challenges and Implications for Intermediate-mass Exoplanets}}.
\newblock \bibinfo{journal}{The Astrophysical Journal} \bibinfo{volume}{789}, \bibinfo{pages}{69}.
\newblock \DOIprefix\doi{10.1088/0004-637X/789/1/69}, \href{http://arxiv.org/abs/1404.5018}{\tt arXiv:1404.5018}.
\bibitem[{{Helled} et~al.(2020){Helled}, {Nettelmann} and {Guillot}}]{helledetal20}
\bibinfo{author}{{Helled}, R.}, \bibinfo{author}{{Nettelmann}, N.}, \bibinfo{author}{{Guillot}, T.}, \bibinfo{year}{2020}.
\newblock \bibinfo{title}{{Uranus and Neptune: Origin, Evolution and Internal Structure}}.
\newblock \bibinfo{journal}{Space Science Reviews} \bibinfo{volume}{216}, \bibinfo{pages}{38}.
\newblock \DOIprefix\doi{10.1007/s11214-020-00660-3}, \href{http://arxiv.org/abs/1909.04891}{\tt arXiv:1909.04891}.
\bibitem[{{Izidoro} et~al.(2021){Izidoro}, {Bitsch}, {Raymond}, {Johansen}, {Morbidelli}, {Lambrechts} and {Jacobson}}]{izidoroetal21}
\bibinfo{author}{{Izidoro}, A.}, \bibinfo{author}{{Bitsch}, B.}, \bibinfo{author}{{Raymond}, S.N.}, \bibinfo{author}{{Johansen}, A.}, \bibinfo{author}{{Morbidelli}, A.}, \bibinfo{author}{{Lambrechts}, M.}, \bibinfo{author}{{Jacobson}, S.A.}, \bibinfo{year}{2021}.
\newblock \bibinfo{title}{{Formation of planetary systems by pebble accretion and migration. Hot super-Earth systems from breaking compact resonant chains}}.
\newblock \bibinfo{journal}{Astronomy \& Astrophysics} \bibinfo{volume}{650}, \bibinfo{pages}{A152}.
\newblock \DOIprefix\doi{10.1051/0004-6361/201935336}, \href{http://arxiv.org/abs/1902.08772}{\tt arXiv:1902.08772}.
\bibitem[{{Izidoro} et~al.(2022){Izidoro}, {Dasgupta}, {Raymond}, {Deienno}, {Bitsch} and {Isella}}]{izidoroetal22b}
\bibinfo{author}{{Izidoro}, A.}, \bibinfo{author}{{Dasgupta}, R.}, \bibinfo{author}{{Raymond}, S.N.}, \bibinfo{author}{{Deienno}, R.}, \bibinfo{author}{{Bitsch}, B.}, \bibinfo{author}{{Isella}, A.}, \bibinfo{year}{2022}.
\newblock \bibinfo{title}{{Planetesimal rings as the cause of the Solar System's planetary architecture}}.
\newblock \bibinfo{journal}{Nature Astronomy} \bibinfo{volume}{6}, \bibinfo{pages}{357--366}.
\newblock \DOIprefix\doi{10.1038/s41550-021-01557-z}, \href{http://arxiv.org/abs/2112.15558}{\tt arXiv:2112.15558}.
\bibitem[{{Izidoro} et~al.(2015a){Izidoro}, {Morbidelli}, {Raymond}, {Hersant} and {Pierens}}]{izidoroetal15}
\bibinfo{author}{{Izidoro}, A.}, \bibinfo{author}{{Morbidelli}, A.}, \bibinfo{author}{{Raymond}, S.N.}, \bibinfo{author}{{Hersant}, F.}, \bibinfo{author}{{Pierens}, A.}, \bibinfo{year}{2015}a.
\newblock \bibinfo{title}{{Accretion of Uranus and Neptune from inward-migrating planetary embryos blocked by Jupiter and Saturn}}.
\newblock \bibinfo{journal}{Astronomy \& Astrophysics} \bibinfo{volume}{582}, \bibinfo{pages}{A99}.
\newblock \DOIprefix\doi{10.1051/0004-6361/201425525}, \href{http://arxiv.org/abs/1506.03029}{\tt arXiv:1506.03029}.
\bibitem[{{Izidoro} et~al.(2015b){Izidoro}, {Raymond}, {Morbidelli}, {Hersant} and {Pierens}}]{izidoroetal15b}
\bibinfo{author}{{Izidoro}, A.}, \bibinfo{author}{{Raymond}, S.N.}, \bibinfo{author}{{Morbidelli}, A.}, \bibinfo{author}{{Hersant}, F.}, \bibinfo{author}{{Pierens}, A.}, \bibinfo{year}{2015}b.
\newblock \bibinfo{title}{{Gas Giant Planets as Dynamical Barriers to Inward-Migrating Super-Earths}}.
\newblock \bibinfo{journal}{The Astrophysical Journall} \bibinfo{volume}{800}, \bibinfo{pages}{L22}.
\newblock \DOIprefix\doi{10.1088/2041-8205/800/2/L22}, \href{http://arxiv.org/abs/1501.06308}{\tt arXiv:1501.06308}.
\bibitem[{{Jakub{\'\i}k} et~al.(2012){Jakub{\'\i}k}, {Morbidelli}, {Neslu{\v{s}}an} and {Brasser}}]{jakubiketal12}
\bibinfo{author}{{Jakub{\'\i}k}, M.}, \bibinfo{author}{{Morbidelli}, A.}, \bibinfo{author}{{Neslu{\v{s}}an}, L.}, \bibinfo{author}{{Brasser}, R.}, \bibinfo{year}{2012}.
\newblock \bibinfo{title}{{The accretion of Uranus and Neptune by collisions among planetary embryos in the vicinity of Jupiter and Saturn}}.
\newblock \bibinfo{journal}{Astronomy \& Astrophysics} \bibinfo{volume}{540}, \bibinfo{pages}{A71}.
\newblock \DOIprefix\doi{10.1051/0004-6361/201117687}, \href{http://arxiv.org/abs/1107.2235}{\tt arXiv:1107.2235}.
\bibitem[{{Johansen} and {Lambrechts}(2017)}]{johansenlambrechts17}
\bibinfo{author}{{Johansen}, A.}, \bibinfo{author}{{Lambrechts}, M.}, \bibinfo{year}{2017}.
\newblock \bibinfo{title}{{Forming Planets via Pebble Accretion}}.
\newblock \bibinfo{journal}{Annual Review of Earth and Planetary Sciences} \bibinfo{volume}{45}, \bibinfo{pages}{359--387}.
\newblock \DOIprefix\doi{10.1146/annurev-earth-063016-020226}.
\bibitem[{{Johansen} et~al.(2015){Johansen}, {Mac Low}, {Lacerda} and {Bizzarro}}]{johansenetal15}
\bibinfo{author}{{Johansen}, A.}, \bibinfo{author}{{Mac Low}, M.M.}, \bibinfo{author}{{Lacerda}, P.}, \bibinfo{author}{{Bizzarro}, M.}, \bibinfo{year}{2015}.
\newblock \bibinfo{title}{{Growth of asteroids, planetary embryos, and Kuiper belt objects by chondrule accretion}}.
\newblock \bibinfo{journal}{Science Advances} \bibinfo{volume}{1}, \bibinfo{pages}{1500109}.
\newblock \DOIprefix\doi{10.1126/sciadv.1500109}, \href{http://arxiv.org/abs/1503.07347}{\tt arXiv:1503.07347}.
\bibitem[{{Johansen} et~al.(2007){Johansen}, {Oishi}, {Mac Low}, {Klahr}, {Henning} and {Youdin}}]{johansenetal07}
\bibinfo{author}{{Johansen}, A.}, \bibinfo{author}{{Oishi}, J.S.}, \bibinfo{author}{{Mac Low}, M.M.}, \bibinfo{author}{{Klahr}, H.}, \bibinfo{author}{{Henning}, T.}, \bibinfo{author}{{Youdin}, A.}, \bibinfo{year}{2007}.
\newblock \bibinfo{title}{{Rapid planetesimal formation in turbulent circumstellar disks}}.
\newblock \bibinfo{journal}{Nature} \bibinfo{volume}{448}, \bibinfo{pages}{1022--1025}.
\newblock \DOIprefix\doi{10.1038/nature06086}, \href{http://arxiv.org/abs/0708.3890}{\tt arXiv:0708.3890}.
\bibitem[{{Kegerreis} et~al.(2018){Kegerreis}, {Teodoro}, {Eke}, {Massey}, {Catling}, {Fryer}, {Korycansky}, {Warren} and {Zahnle}}]{kegerreisetal18}
\bibinfo{author}{{Kegerreis}, J.A.}, \bibinfo{author}{{Teodoro}, L.F.A.}, \bibinfo{author}{{Eke}, V.R.}, \bibinfo{author}{{Massey}, R.J.}, \bibinfo{author}{{Catling}, D.C.}, \bibinfo{author}{{Fryer}, C.L.}, \bibinfo{author}{{Korycansky}, D.G.}, \bibinfo{author}{{Warren}, M.S.}, \bibinfo{author}{{Zahnle}, K.J.}, \bibinfo{year}{2018}.
\newblock \bibinfo{title}{{Consequences of Giant Impacts on Early Uranus for Rotation, Internal Structure, Debris, and Atmospheric Erosion}}.
\newblock \bibinfo{journal}{The Astrophysical Journal} \bibinfo{volume}{861}, \bibinfo{pages}{52}.
\newblock \DOIprefix\doi{10.3847/1538-4357/aac725}, \href{http://arxiv.org/abs/1803.07083}{\tt arXiv:1803.07083}.
\bibitem[{{Kokubo} and {Ida}(1998)}]{kokuboida98}
\bibinfo{author}{{Kokubo}, E.}, \bibinfo{author}{{Ida}, S.}, \bibinfo{year}{1998}.
\newblock \bibinfo{title}{{Oligarchic Growth of Protoplanets}}.
\newblock \bibinfo{journal}{Icarus} \bibinfo{volume}{131}, \bibinfo{pages}{171--178}.
\newblock \DOIprefix\doi{10.1006/icar.1997.5840}.
\bibitem[{{Kokubo} and {Ida}(2007)}]{kokuboetal07}
\bibinfo{author}{{Kokubo}, E.}, \bibinfo{author}{{Ida}, S.}, \bibinfo{year}{2007}.
\newblock \bibinfo{title}{{Formation of Terrestrial Planets from Protoplanets. II. Statistics of Planetary Spin}}.
\newblock \bibinfo{journal}{The Astrophysical Journal} \bibinfo{volume}{671}, \bibinfo{pages}{2082--2090}.
\newblock \DOIprefix\doi{10.1086/522364}.
\bibitem[{{Kurosaki} and {Inutsuka}(2019)}]{kurosakiinutsuka19}
\bibinfo{author}{{Kurosaki}, K.}, \bibinfo{author}{{Inutsuka}, S.i.}, \bibinfo{year}{2019}.
\newblock \bibinfo{title}{{The Exchange of Mass and Angular Momentum in the Impact Event of Ice Giant Planets: Implications for the Origin of Uranus}}.
\newblock \bibinfo{journal}{The Astronomical Journal} \bibinfo{volume}{157}, \bibinfo{pages}{13}.
\newblock \DOIprefix\doi{10.3847/1538-3881/aaf165}, \href{http://arxiv.org/abs/1811.05234}{\tt arXiv:1811.05234}.
\bibitem[{{Lambrechts} and {Johansen}(2012)}]{lambrechtsetal12}
\bibinfo{author}{{Lambrechts}, M.}, \bibinfo{author}{{Johansen}, A.}, \bibinfo{year}{2012}.
\newblock \bibinfo{title}{{Rapid growth of gas-giant cores by pebble accretion}}.
\newblock \bibinfo{journal}{Astronomy \& Astrophysics} \bibinfo{volume}{544}, \bibinfo{pages}{A32}.
\newblock \DOIprefix\doi{10.1051/0004-6361/201219127}, \href{http://arxiv.org/abs/1205.3030}{\tt arXiv:1205.3030}.
\bibitem[{{Lambrechts} et~al.(2019){Lambrechts}, {Morbidelli}, {Jacobson}, {Johansen}, {Bitsch}, {Izidoro} and {Raymond}}]{lambrechtsetal19}
\bibinfo{author}{{Lambrechts}, M.}, \bibinfo{author}{{Morbidelli}, A.}, \bibinfo{author}{{Jacobson}, S.A.}, \bibinfo{author}{{Johansen}, A.}, \bibinfo{author}{{Bitsch}, B.}, \bibinfo{author}{{Izidoro}, A.}, \bibinfo{author}{{Raymond}, S.N.}, \bibinfo{year}{2019}.
\newblock \bibinfo{title}{{Formation of planetary systems by pebble accretion and migration. How the radial pebble flux determines a terrestrial-planet or super-Earth growth mode}}.
\newblock \bibinfo{journal}{Astronomy \& Astrophysics} \bibinfo{volume}{627}, \bibinfo{pages}{A83}.
\newblock \DOIprefix\doi{10.1051/0004-6361/201834229}, \href{http://arxiv.org/abs/1902.08694}{\tt arXiv:1902.08694}.
\bibitem[{{Lau} et~al.(2022){Lau}, {Dr{\k{a}}{\.z}kowska}, {Stammler}, {Birnstiel} and {Dullemond}}]{lauetal22}
\bibinfo{author}{{Lau}, T.C.H.}, \bibinfo{author}{{Dr{\k{a}}{\.z}kowska}, J.}, \bibinfo{author}{{Stammler}, S.M.}, \bibinfo{author}{{Birnstiel}, T.}, \bibinfo{author}{{Dullemond}, C.P.}, \bibinfo{year}{2022}.
\newblock \bibinfo{title}{{Rapid formation of massive planetary cores in a pressure bump}}.
\newblock \bibinfo{journal}{Astronomy \& Astrophysics} \bibinfo{volume}{668}, \bibinfo{pages}{A170}.
\newblock \DOIprefix\doi{10.1051/0004-6361/202244864}, \href{http://arxiv.org/abs/2211.04497}{\tt arXiv:2211.04497}.
\bibitem[{{Lau} et~al.(2024){Lau}, {Lee}, {Brasser} and {Matsumura}}]{lauetal24}
\bibinfo{author}{{Lau}, T.C.H.}, \bibinfo{author}{{Lee}, M.H.}, \bibinfo{author}{{Brasser}, R.}, \bibinfo{author}{{Matsumura}, S.}, \bibinfo{year}{2024}.
\newblock \bibinfo{title}{{Can the giant planets of the Solar System form via pebble accretion in a smooth protoplanetary disc?}}
\newblock \bibinfo{journal}{Astronomy \& Astrophysics} \bibinfo{volume}{683}, \bibinfo{pages}{A204}.
\newblock \DOIprefix\doi{10.1051/0004-6361/202347863}, \href{http://arxiv.org/abs/2401.05036}{\tt arXiv:2401.05036}.
\bibitem[{{Levison} et~al.(2015){Levison}, {Kretke} and {Duncan}}]{levisonetal15}
\bibinfo{author}{{Levison}, H.F.}, \bibinfo{author}{{Kretke}, K.A.}, \bibinfo{author}{{Duncan}, M.J.}, \bibinfo{year}{2015}.
\newblock \bibinfo{title}{{Growing the gas-giant planets by the gradual accumulation of pebbles}}.
\newblock \bibinfo{journal}{Nature} \bibinfo{volume}{524}, \bibinfo{pages}{322--324}.
\newblock \DOIprefix\doi{10.1038/nature14675}, \href{http://arxiv.org/abs/1510.02094}{\tt arXiv:1510.02094}.
\bibitem[{{Levison} and {Morbidelli}(2007)}]{levisonmorbidelli07}
\bibinfo{author}{{Levison}, H.F.}, \bibinfo{author}{{Morbidelli}, A.}, \bibinfo{year}{2007}.
\newblock \bibinfo{title}{{Models of the collisional damping scenario for ice-giant planets and Kuiper belt formation}}.
\newblock \bibinfo{journal}{Icarus} \bibinfo{volume}{189}, \bibinfo{pages}{196--212}.
\newblock \DOIprefix\doi{10.1016/j.icarus.2007.01.004}, \href{http://arxiv.org/abs/astro-ph/0701544}{\tt arXiv:astro-ph/0701544}.
\bibitem[{{Levison} and {Stewart}(2001)}]{levisonstewart01}
\bibinfo{author}{{Levison}, H.F.}, \bibinfo{author}{{Stewart}, G.R.}, \bibinfo{year}{2001}.
\newblock \bibinfo{title}{{Remarks on Modeling the Formation of Uranus and Neptune}}.
\newblock \bibinfo{journal}{Icarus} \bibinfo{volume}{153}, \bibinfo{pages}{224--228}.
\newblock \DOIprefix\doi{10.1006/icar.2001.6672}.
\bibitem[{{Levison} et~al.(2010){Levison}, {Thommes} and {Duncan}}]{levisonetal10}
\bibinfo{author}{{Levison}, H.F.}, \bibinfo{author}{{Thommes}, E.}, \bibinfo{author}{{Duncan}, M.J.}, \bibinfo{year}{2010}.
\newblock \bibinfo{title}{{Modeling the Formation of Giant Planet Cores. I. Evaluating Key Processes}}.
\newblock \bibinfo{journal}{The Astronomical Journal} \bibinfo{volume}{139}, \bibinfo{pages}{1297--1314}.
\newblock \DOIprefix\doi{10.1088/0004-6256/139/4/1297}, \href{http://arxiv.org/abs/0912.3144}{\tt arXiv:0912.3144}.
\bibitem[{{Lyra} et~al.(2008){Lyra}, {Johansen}, {Klahr} and {Piskunov}}]{lyraetal08}
\bibinfo{author}{{Lyra}, W.}, \bibinfo{author}{{Johansen}, A.}, \bibinfo{author}{{Klahr}, H.}, \bibinfo{author}{{Piskunov}, N.}, \bibinfo{year}{2008}.
\newblock \bibinfo{title}{{Embryos grown in the dead zone. Assembling the first protoplanetary cores in low mass self-gravitating circumstellar disks of gas and solids}}.
\newblock \bibinfo{journal}{Astronomy \& Astrophysics} \bibinfo{volume}{491}, \bibinfo{pages}{L41--L44}.
\newblock \DOIprefix\doi{10.1051/0004-6361:200810626}, \href{http://arxiv.org/abs/0807.2622}{\tt arXiv:0807.2622}.
\bibitem[{{Malhotra}(1993)}]{malhotra93}
\bibinfo{author}{{Malhotra}, R.}, \bibinfo{year}{1993}.
\newblock \bibinfo{title}{{The origin of Pluto's peculiar orbit}}.
\newblock \bibinfo{journal}{Nature} \bibinfo{volume}{365}, \bibinfo{pages}{819--821}.
\newblock \DOIprefix\doi{10.1038/365819a0}.
\bibitem[{{Masset} et~al.(2006){Masset}, {Morbidelli}, {Crida} and {Ferreira}}]{massetetal06}
\bibinfo{author}{{Masset}, F.S.}, \bibinfo{author}{{Morbidelli}, A.}, \bibinfo{author}{{Crida}, A.}, \bibinfo{author}{{Ferreira}, J.}, \bibinfo{year}{2006}.
\newblock \bibinfo{title}{{Disk Surface Density Transitions as Protoplanet Traps}}.
\newblock \bibinfo{journal}{The Astrophysical Journal} \bibinfo{volume}{642}, \bibinfo{pages}{478--487}.
\newblock \DOIprefix\doi{10.1086/500967}.
\bibitem[{{Morbidelli}(2020)}]{morbidelli20}
\bibinfo{author}{{Morbidelli}, A.}, \bibinfo{year}{2020}.
\newblock \bibinfo{title}{{Planet formation by pebble accretion in ringed disks}}.
\newblock \bibinfo{journal}{Astronomy \& Astrophysics} \bibinfo{volume}{638}, \bibinfo{pages}{A1}.
\newblock \DOIprefix\doi{10.1051/0004-6361/202037983}, \href{http://arxiv.org/abs/2004.04942}{\tt arXiv:2004.04942}.
\bibitem[{{Morbidelli} and {Crida}(2007)}]{morbidellicrida07}
\bibinfo{author}{{Morbidelli}, A.}, \bibinfo{author}{{Crida}, A.}, \bibinfo{year}{2007}.
\newblock \bibinfo{title}{{The dynamics of Jupiter and Saturn in the gaseous protoplanetary disk}}.
\newblock \bibinfo{journal}{Icarus} \bibinfo{volume}{191}, \bibinfo{pages}{158--171}.
\newblock \DOIprefix\doi{10.1016/j.icarus.2007.04.001}.
\bibitem[{{Morbidelli} et~al.(2012){Morbidelli}, {Tsiganis}, {Batygin}, {Crida} and {Gomes}}]{morbidellietal12}
\bibinfo{author}{{Morbidelli}, A.}, \bibinfo{author}{{Tsiganis}, K.}, \bibinfo{author}{{Batygin}, K.}, \bibinfo{author}{{Crida}, A.}, \bibinfo{author}{{Gomes}, R.}, \bibinfo{year}{2012}.
\newblock \bibinfo{title}{{Explaining why the uranian satellites have equatorial prograde orbits despite the large planetary obliquity}}.
\newblock \bibinfo{journal}{Icarus} \bibinfo{volume}{219}, \bibinfo{pages}{737--740}.
\newblock \DOIprefix\doi{10.1016/j.icarus.2012.03.025}, \href{http://arxiv.org/abs/1208.4685}{\tt arXiv:1208.4685}.
\bibitem[{{Nakagawa} et~al.(1986){Nakagawa}, {Sekiya} and {Hayashi}}]{nakagawaetal86}
\bibinfo{author}{{Nakagawa}, Y.}, \bibinfo{author}{{Sekiya}, M.}, \bibinfo{author}{{Hayashi}, C.}, \bibinfo{year}{1986}.
\newblock \bibinfo{title}{{Settling and growth of dust particles in a laminar phase of a low-mass solar nebula}}.
\newblock \bibinfo{journal}{Icarus} \bibinfo{volume}{67}, \bibinfo{pages}{375--390}.
\newblock \DOIprefix\doi{10.1016/0019-1035(86)90121-1}.
\bibitem[{{Nesvorn{\'y}}(2011)}]{nesvorny11}
\bibinfo{author}{{Nesvorn{\'y}}, D.}, \bibinfo{year}{2011}.
\newblock \bibinfo{title}{{Young Solar System's Fifth Giant Planet?}}
\newblock \bibinfo{journal}{The Astrophysical Journall} \bibinfo{volume}{742}, \bibinfo{pages}{L22}.
\newblock \DOIprefix\doi{10.1088/2041-8205/742/2/L22}, \href{http://arxiv.org/abs/1109.2949}{\tt arXiv:1109.2949}.
\bibitem[{{Nesvorn{\'y}} and {Morbidelli}(2012)}]{nesvornymorbidelli12}
\bibinfo{author}{{Nesvorn{\'y}}, D.}, \bibinfo{author}{{Morbidelli}, A.}, \bibinfo{year}{2012}.
\newblock \bibinfo{title}{{Statistical Study of the Early Solar System's Instability with Four, Five, and Six Giant Planets}}.
\newblock \bibinfo{journal}{The Astronomical Journal} \bibinfo{volume}{144}, \bibinfo{pages}{117}.
\newblock \DOIprefix\doi{10.1088/0004-6256/144/4/117}, \href{http://arxiv.org/abs/1208.2957}{\tt arXiv:1208.2957}.
\bibitem[{{Nettelmann} et~al.(2013){Nettelmann}, {Helled}, {Fortney} and {Redmer}}]{nettelmannetal13}
\bibinfo{author}{{Nettelmann}, N.}, \bibinfo{author}{{Helled}, R.}, \bibinfo{author}{{Fortney}, J.J.}, \bibinfo{author}{{Redmer}, R.}, \bibinfo{year}{2013}.
\newblock \bibinfo{title}{{New indication for a dichotomy in the interior structure of Uranus and Neptune from the application of modified shape and rotation data}}.
\newblock \bibinfo{journal}{Planetary and Space Science} \bibinfo{volume}{77}, \bibinfo{pages}{143--151}.
\newblock \DOIprefix\doi{10.1016/j.pss.2012.06.019}, \href{http://arxiv.org/abs/1207.2309}{\tt arXiv:1207.2309}.
\bibitem[{{Ormel} and {Klahr}(2010)}]{ormelklahr10}
\bibinfo{author}{{Ormel}, C.W.}, \bibinfo{author}{{Klahr}, H.H.}, \bibinfo{year}{2010}.
\newblock \bibinfo{title}{{The effect of gas drag on the growth of protoplanets. Analytical expressions for the accretion of small bodies in laminar disks}}.
\newblock \bibinfo{journal}{Astronomy \& Astrophysics} \bibinfo{volume}{520}, \bibinfo{pages}{A43}.
\newblock \DOIprefix\doi{10.1051/0004-6361/201014903}, \href{http://arxiv.org/abs/1007.0916}{\tt arXiv:1007.0916}.
\bibitem[{{Paardekooper} et~al.(2011){Paardekooper}, {Baruteau} and {Kley}}]{paardekooperetal11}
\bibinfo{author}{{Paardekooper}, S.J.}, \bibinfo{author}{{Baruteau}, C.}, \bibinfo{author}{{Kley}, W.}, \bibinfo{year}{2011}.
\newblock \bibinfo{title}{{A torque formula for non-isothermal Type I planetary migration - II. Effects of diffusion}}.
\newblock \bibinfo{journal}{Monthly Notices of the Royal Astronomical Society} \bibinfo{volume}{410}, \bibinfo{pages}{293--303}.
\newblock \DOIprefix\doi{10.1111/j.1365-2966.2010.17442.x}, \href{http://arxiv.org/abs/1007.4964}{\tt arXiv:1007.4964}.
\bibitem[{{Pirani} et~al.(2021){Pirani}, {Johansen} and {Mustill}}]{piranietal21}
\bibinfo{author}{{Pirani}, S.}, \bibinfo{author}{{Johansen}, A.}, \bibinfo{author}{{Mustill}, A.J.}, \bibinfo{year}{2021}.
\newblock \bibinfo{title}{{How the formation of Neptune shapes the Kuiper belt}}.
\newblock \bibinfo{journal}{Astronomy \& Astrophysics} \bibinfo{volume}{650}, \bibinfo{pages}{A161}.
\newblock \DOIprefix\doi{10.1051/0004-6361/202037465}, \href{http://arxiv.org/abs/2104.12267}{\tt arXiv:2104.12267}.
\bibitem[{{Podolak} and {Helled}(2012)}]{podolakhelled12}
\bibinfo{author}{{Podolak}, M.}, \bibinfo{author}{{Helled}, R.}, \bibinfo{year}{2012}.
\newblock \bibinfo{title}{{What Do We Really Know about Uranus and Neptune?}}
\newblock \bibinfo{journal}{The Astrophysical Journall} \bibinfo{volume}{759}, \bibinfo{pages}{L32}.
\newblock \DOIprefix\doi{10.1088/2041-8205/759/2/L32}, \href{http://arxiv.org/abs/1208.5551}{\tt arXiv:1208.5551}.
\bibitem[{{Pollack} et~al.(1996){Pollack}, {Hubickyj}, {Bodenheimer}, {Lissauer}, {Podolak} and {Greenzweig}}]{pollacketal96}
\bibinfo{author}{{Pollack}, J.B.}, \bibinfo{author}{{Hubickyj}, O.}, \bibinfo{author}{{Bodenheimer}, P.}, \bibinfo{author}{{Lissauer}, J.J.}, \bibinfo{author}{{Podolak}, M.}, \bibinfo{author}{{Greenzweig}, Y.}, \bibinfo{year}{1996}.
\newblock \bibinfo{title}{{Formation of the Giant Planets by Concurrent Accretion of Solids and Gas}}.
\newblock \bibinfo{journal}{Icarus} \bibinfo{volume}{124}, \bibinfo{pages}{62--85}.
\newblock \DOIprefix\doi{10.1006/icar.1996.0190}.
\bibitem[{{Reinhardt} et~al.(2020){Reinhardt}, {Chau}, {Stadel} and {Helled}}]{reinhardtetal20}
\bibinfo{author}{{Reinhardt}, C.}, \bibinfo{author}{{Chau}, A.}, \bibinfo{author}{{Stadel}, J.}, \bibinfo{author}{{Helled}, R.}, \bibinfo{year}{2020}.
\newblock \bibinfo{title}{{Bifurcation in the history of Uranus and Neptune: the role of giant impacts}}.
\newblock \bibinfo{journal}{Monthly Notices of the Royal Astronomical Society} \bibinfo{volume}{492}, \bibinfo{pages}{5336--5353}.
\newblock \DOIprefix\doi{10.1093/mnras/stz3271}, \href{http://arxiv.org/abs/1907.09809}{\tt arXiv:1907.09809}.
\bibitem[{{Rogoszinski} and {Hamilton}(2021)}]{rogoszinskihamilton21}
\bibinfo{author}{{Rogoszinski}, Z.}, \bibinfo{author}{{Hamilton}, D.P.}, \bibinfo{year}{2021}.
\newblock \bibinfo{title}{{Tilting Uranus: Collisions versus Spin-Orbit Resonance}}.
\newblock \bibinfo{journal}{The Planetary Science Journal} \bibinfo{volume}{2}, \bibinfo{pages}{78}.
\newblock \DOIprefix\doi{10.3847/PSJ/abec4e}, \href{http://arxiv.org/abs/2004.14913}{\tt arXiv:2004.14913}.
\bibitem[{{Rufu} and {Canup}(2022)}]{rufucanup22}
\bibinfo{author}{{Rufu}, R.}, \bibinfo{author}{{Canup}, R.M.}, \bibinfo{year}{2022}.
\newblock \bibinfo{title}{{Coaccretion + Giant-impact Origin of the Uranus System: Tilting Impact}}.
\newblock \bibinfo{journal}{The Astrophysical Journal} \bibinfo{volume}{928}, \bibinfo{pages}{123}.
\newblock \DOIprefix\doi{10.3847/1538-4357/ac525a}, \href{http://arxiv.org/abs/2204.00124}{\tt arXiv:2204.00124}.
\bibitem[{{Safronov}(1972)}]{safronov72}
\bibinfo{author}{{Safronov}, V.S.}, \bibinfo{year}{1972}.
\newblock \bibinfo{title}{{Evolution of the protoplanetary cloud and formation of the earth and planets.}}
\bibitem[{{Salmon} and {Canup}(2022)}]{salmoncanup22}
\bibinfo{author}{{Salmon}, J.}, \bibinfo{author}{{Canup}, R.M.}, \bibinfo{year}{2022}.
\newblock \bibinfo{title}{{Co-accretion + Giant Impact Origin of the Uranus System: Post-impact Evolution}}.
\newblock \bibinfo{journal}{The Astrophysical Journal} \bibinfo{volume}{924}, \bibinfo{pages}{6}.
\newblock \DOIprefix\doi{10.3847/1538-4357/ac300e}.
\bibitem[{{S{\'a}ndor} et~al.(2024){S{\'a}ndor}, {Guilera}, {Reg{\'a}ly} and {Lyra}}]{sandoretal24}
\bibinfo{author}{{S{\'a}ndor}, Z.}, \bibinfo{author}{{Guilera}, O.M.}, \bibinfo{author}{{Reg{\'a}ly}, Z.}, \bibinfo{author}{{Lyra}, W.}, \bibinfo{year}{2024}.
\newblock \bibinfo{title}{{Planetesimal and planet formation in transient dust traps}}.
\newblock \bibinfo{journal}{Astronomy \& Astrophysics} \bibinfo{volume}{686}, \bibinfo{pages}{A78}.
\newblock \DOIprefix\doi{10.1051/0004-6361/202347605}, \href{http://arxiv.org/abs/2402.11584}{\tt arXiv:2402.11584}.
\bibitem[{{Tanaka} and {Ida}(1997)}]{tanakaida97}
\bibinfo{author}{{Tanaka}, H.}, \bibinfo{author}{{Ida}, S.}, \bibinfo{year}{1997}.
\newblock \bibinfo{title}{{Distribution of Planetesimals around a Protoplanet in the Nebula Gas}}.
\newblock \bibinfo{journal}{Icarus} \bibinfo{volume}{125}, \bibinfo{pages}{302--316}.
\newblock \DOIprefix\doi{10.1006/icar.1996.9998}.
\bibitem[{{Tanaka} et~al.(2002){Tanaka}, {Takeuchi} and {Ward}}]{tanakaetal02}
\bibinfo{author}{{Tanaka}, H.}, \bibinfo{author}{{Takeuchi}, T.}, \bibinfo{author}{{Ward}, W.R.}, \bibinfo{year}{2002}.
\newblock \bibinfo{title}{{Three-Dimensional Interaction between a Planet and an Isothermal Gaseous Disk. I. Corotation and Lindblad Torques and Planet Migration}}.
\newblock \bibinfo{journal}{The Astrophysical Journal} \bibinfo{volume}{565}, \bibinfo{pages}{1257--1274}.
\newblock \DOIprefix\doi{10.1086/324713}.
\bibitem[{{Tanaka} and {Ward}(2004)}]{tanakaward04}
\bibinfo{author}{{Tanaka}, H.}, \bibinfo{author}{{Ward}, W.R.}, \bibinfo{year}{2004}.
\newblock \bibinfo{title}{{Three-dimensional Interaction between a Planet and an Isothermal Gaseous Disk. II. Eccentricity Waves and Bending Waves}}.
\newblock \bibinfo{journal}{The Astrophysical Journal} \bibinfo{volume}{602}, \bibinfo{pages}{388--395}.
\newblock \DOIprefix\doi{10.1086/380992}.
\bibitem[{{Thommes} et~al.(2003){Thommes}, {Duncan} and {Levison}}]{thommesetal03}
\bibinfo{author}{{Thommes}, E.W.}, \bibinfo{author}{{Duncan}, M.J.}, \bibinfo{author}{{Levison}, H.F.}, \bibinfo{year}{2003}.
\newblock \bibinfo{title}{{Oligarchic growth of giant planets}}.
\newblock \bibinfo{journal}{Icarus} \bibinfo{volume}{161}, \bibinfo{pages}{431--455}.
\newblock \DOIprefix\doi{10.1016/S0019-1035(02)00043-X}, \href{http://arxiv.org/abs/astro-ph/0303269}{\tt arXiv:astro-ph/0303269}.
\bibitem[{{Tsiganis} et~al.(2005){Tsiganis}, {Gomes}, {Morbidelli} and {Levison}}]{tsiganisetal05}
\bibinfo{author}{{Tsiganis}, K.}, \bibinfo{author}{{Gomes}, R.}, \bibinfo{author}{{Morbidelli}, A.}, \bibinfo{author}{{Levison}, H.F.}, \bibinfo{year}{2005}.
\newblock \bibinfo{title}{{Origin of the orbital architecture of the giant planets of the Solar System}}.
\newblock \bibinfo{journal}{Nature} \bibinfo{volume}{435}, \bibinfo{pages}{459--461}.
\newblock \DOIprefix\doi{10.1038/nature03539}.
\bibitem[{{Valletta} and {Helled}(2022)}]{vallettahelled22}
\bibinfo{author}{{Valletta}, C.}, \bibinfo{author}{{Helled}, R.}, \bibinfo{year}{2022}.
\newblock \bibinfo{title}{{Possible In Situ Formation of Uranus and Neptune via Pebble Accretion}}.
\newblock \bibinfo{journal}{The Astrophysical Journal} \bibinfo{volume}{931}, \bibinfo{pages}{21}.
\newblock \DOIprefix\doi{10.3847/1538-4357/ac5f52}, \href{http://arxiv.org/abs/2203.06545}{\tt arXiv:2203.06545}.
\bibitem[{{Visser} et~al.(2020){Visser}, {Ormel}, {Dominik} and {Ida}}]{visseretal20}
\bibinfo{author}{{Visser}, R.G.}, \bibinfo{author}{{Ormel}, C.W.}, \bibinfo{author}{{Dominik}, C.}, \bibinfo{author}{{Ida}, S.}, \bibinfo{year}{2020}.
\newblock \bibinfo{title}{{Spinning up planetary bodies by pebble accretion}}.
\newblock \bibinfo{journal}{Icarus} \bibinfo{volume}{335}, \bibinfo{pages}{113380}.
\newblock \DOIprefix\doi{10.1016/j.icarus.2019.07.014}, \href{http://arxiv.org/abs/1907.04368}{\tt arXiv:1907.04368}.
\bibitem[{{Walsh} and {Morbidelli}(2011)}]{walshmorbidelli11}
\bibinfo{author}{{Walsh}, K.J.}, \bibinfo{author}{{Morbidelli}, A.}, \bibinfo{year}{2011}.
\newblock \bibinfo{title}{{The effect of an early planetesimal-driven migration of the giant planets on terrestrial planet formation}}.
\newblock \bibinfo{journal}{Astronomy and Astrophysics} \bibinfo{volume}{526}, \bibinfo{pages}{A126}.
\newblock \DOIprefix\doi{10.1051/0004-6361/201015277}, \href{http://arxiv.org/abs/1101.3776}{\tt arXiv:1101.3776}.
\bibitem[{{Ward}(1986)}]{ward86}
\bibinfo{author}{{Ward}, W.R.}, \bibinfo{year}{1986}.
\newblock \bibinfo{title}{{Density waves in the solar nebula: Diffential Lindblad torque}}.
\newblock \bibinfo{journal}{Icarus} \bibinfo{volume}{67}, \bibinfo{pages}{164--180}.
\newblock \DOIprefix\doi{10.1016/0019-1035(86)90182-X}.
\bibitem[{{Wetherill} and {Stewart}(1989)}]{wetherillstewart89}
\bibinfo{author}{{Wetherill}, G.W.}, \bibinfo{author}{{Stewart}, G.R.}, \bibinfo{year}{1989}.
\newblock \bibinfo{title}{{Accumulation of a swarm of small planetesimals}}.
\newblock \bibinfo{journal}{Icarus} \bibinfo{volume}{77}, \bibinfo{pages}{330--357}.
\newblock \DOIprefix\doi{10.1016/0019-1035(89)90093-6}.
\bibitem[{{Woo} et~al.(2022){Woo}, {Reinhardt}, {Cilibrasi}, {Chau}, {Helled} and {Stadel}}]{wooetal2022}
\bibinfo{author}{{Woo}, J.M.Y.}, \bibinfo{author}{{Reinhardt}, C.}, \bibinfo{author}{{Cilibrasi}, M.}, \bibinfo{author}{{Chau}, A.}, \bibinfo{author}{{Helled}, R.}, \bibinfo{author}{{Stadel}, J.}, \bibinfo{year}{2022}.
\newblock \bibinfo{title}{{Did Uranus' regular moons form via a rocky giant impactor?}}
\newblock \bibinfo{journal}{Icarus} \bibinfo{volume}{375}, \bibinfo{pages}{114842}.
\newblock \DOIprefix\doi{10.1016/j.icarus.2021.114842}, \href{http://arxiv.org/abs/2105.13663}{\tt arXiv:2105.13663}.
\bibitem[{{Youdin} and {Goodman}(2005)}]{youdingoodman05}
\bibinfo{author}{{Youdin}, A.N.}, \bibinfo{author}{{Goodman}, J.}, \bibinfo{year}{2005}.
\newblock \bibinfo{title}{{Streaming Instabilities in Protoplanetary Disks}}.
\newblock \bibinfo{journal}{The Astrophysical Journal} \bibinfo{volume}{620}, \bibinfo{pages}{459--469}.
\newblock \DOIprefix\doi{10.1086/426895}, \href{http://arxiv.org/abs/astro-ph/0409263}{\tt arXiv:astro-ph/0409263}.

\end{thebibliography}




\end{document}